\documentclass[preprint,aps,prd,amsmath,amssymb]{revtex4}
\pdfoutput=1
\usepackage{graphicx}% Include figure files
\usepackage{dcolumn}% Align table columns on decimal point
\usepackage{bm}% bold math
\usepackage{amsmath}
\usepackage{amsfonts}
\usepackage{bbm}
\usepackage{subfigure}
\usepackage{setspace}
\usepackage{color}
\usepackage{pxfonts}

\newcommand{\beq}{\begin{equation}}
\newcommand{\eeq}{\end{equation}}
\newcommand{\bea}{\begin{eqnarray}}
\newcommand{\eea}{\end{eqnarray}}
\newcommand{\ba}{\begin{array}}
\newcommand{\ea}{\end{array}}
\newcommand{\bi}{\begin{itemize}}
\newcommand{\ei}{\end{itemize}}
\newcommand{\bn}{\begin{enumerate}}
\newcommand{\en}{\end{enumerate}}
\newcommand{\bc}{\begin{center}}
\newcommand{\ec}{\end{center}}

\newcommand{\no}{\nonumber}
\newcommand{\nn}{\nonumber\\}

\newcommand{\gsim}{\lower.7ex\hbox{$\;\stackrel{\textstyle>}{\sim}\;$}}
\newcommand{\lsim}{\lower.7ex\hbox{$\;\stackrel{\textstyle<}{\sim}\;$}}

\def\met{{\slash\!\!\!\!\!\:E}_T}
\def\mpt{{\slash\!\!\!\!\!\:p}_T}

\begin{document}

%%%%%%%%%%%%%%%%%%%%%%%%%%%%%%%%%%%%%%%%%%%%%%%%%%%

\title{\Large \color{red} Fourth Lepton Family is Natural in Technicolor}
\author{Mads T. {\sc Frandsen} $^{\color{blue}{\varheartsuit}\color{blue}{\spadesuit}}$ }
\email{toudal@ifk.sdu.dk}
\author{Isabella {\sc Masina}$^{\color{blue}{\clubsuit} \color{blue}{\varheartsuit}}$}
\email{masina@fe.infn.it}
\author{Francesco {\sc Sannino}$^{\color{blue}{\varheartsuit}}$}
\email{sannino@ifk.sdu.dk}
\affiliation{$^{\color{blue}{\varheartsuit}}$  CP$^\mathbf 3$ - Origins, IFK \& IMADA, University of Southern Denmark, Campusvej 55, DK-5230 Odense M, Denmark \footnote{Centre of Excellence for Particle Physics Phenomenology dedicated to the understanding of the  Origins of Mass in the universe. This is the new affiliation from September 1st 2009.} \\
$^{\color{blue}{\spadesuit}}$ {Rudolf Peierls Centre for Theoretical Physics, 
University of Oxford, 
1 Keble Road, Oxford OX1 3NP, United Kingdom.}
\\$^{\color{blue}{\clubsuit}}$ Dip.~di Fisica dell'Universit\`a degli Studi di Ferrara
and INFN Sez.~di Ferrara, Via Saragat 1, I-44100 Ferrara, Italy  }

\begin{abstract}

Imagine to discover a new fourth family of leptons  at the Large Hadron Collider (LHC) but no signs of an associated fourth family of quarks. What would that imply? An intriguing possibility is that the new fermions needed to compensate for the new leptons gauge anomalies simultaneously address the big hierarchy problem of the Standard Model. A natural way to accomplish such a scenario is to have the Higgs itself be composite of these new fermions. This is the setup we are going to investigate in this paper using as a template Minimal Walking Technicolor. We analyze a general heavy neutrino mass structure with and without mixing with the Standard Model families. We also analyze the LHC potential to observe the fourth lepton family in tandem with the new composite Higgs dynamics. 
We finally introduce a model uniting the fourth lepton family and the technifermion sector at higher energies. 

\end{abstract}

%%%%%%%%%%%%%%%%%%%%%%%%%%%%%%%%%%%%%%%%%%%%%%%%%%%%%%%%%%%%%%%%%%%%%%%%

%%%%%%%%%%%%%%%%%%%%%%%%%%%%%%%%%%%%%%%%%%%%%%%%%%%%%%%%%%%%%%%%%%%%%%%%
\maketitle
\newpage
\tableofcontents
\newpage

%%%%%%%%%%%%%%%%%%%%%%%%%%%%%%%%%%%%%%%
\section{Introduction } 

If the LHC discovers a new fourth family of leptons but no associated quarks, what would that imply? Either the associated quarks are much heavier than the electroweak scale or a new set of fermions are needed to account for the 4th family lepton induced gauge anomalies.  The new fermions could address directly the big hierarchy problem of the SM if their dynamics leads to a composite Higgs scenario. This is the setup we are going to investigate in this paper using as a template Minimal Walking Technicolor (MWT)  \cite{Sannino:2004qp}. 

To investigate the phenomenology of such a theory we must first discuss the electromagnetic neutral lepton sector. We will consider a general mass structure for the fourth neutrino and both the case of mixing and no mixing with the SM neutrinos. We will also summarize the (in)direct phenomenological constraints. We then analyze the interplay of the composite Higgs sector with the new lepton family at the LHC.  

We will study the production and decay of the new leptons in proton - proton collisions which is relevant to select the LHC signatures for the discovery of these new leptons. We show that the composite Higgs structure can affect and differentiate the final signatures with respect to the case in which the Higgs is elementary. The bottom line is that one can experimentally determine if the fourth family is associated to a composite Higgs sector.

We then move on to a more general framework in which the fourth family of leptons, together with the technicolor sector unites in a second $SU(2)\times U(1)$ gauge group. 
This model has a number of interesting features and reduces to the one investigated above when the mass scale of the new gauge bosons is sufficiently larger than the electroweak one.

%%%%%%%%%%%%%%%%%%%%%%%%%%%%%%%%%%%%%%%%%%%%%%%%%%%%%%%%%%%%%%%%%%%%%%%%%%%
\section{A natural fourth family of leptons at the TeV-scale}

In the SM, the three lepton families, $\ell=e,\mu,\tau$, belong to the following representations 
of the gauge group $SU(3)_c\times SU(2)_L \times U(1)_Y$:
\beq
L_\ell = ({\nu_\ell}_L ~~ \ell_L)^T \sim (1,2,-1/2) ~~~~~,~~~~~\ell_R \sim (1,1,-1)~~~~~,
\eeq  
where the chirality projectors $P_L=(1-\gamma_5)/2$ and $P_R=(1+\gamma_5)/2$ have been introduced and 
the relation $Q=T_3+Y$ has been adopted in order to define the hypercharge with respect to the electric charge.
The neutral and charged current interactions of the SM leptons are then respectively accounted for by 
the Lagrangian terms: 
\bea
{\cal L}_{NC} &=& 
\frac{g }{\cos \theta_W}  ~\left( \frac{1}{2} \bar \nu_L \gamma^\mu \nu_L - \frac{1}{2} \bar \ell_L \gamma^\mu \ell_L
+ \sin^2 \theta_W ~\bar \ell \gamma^\mu \ell \right) ~Z_\mu ~+ ~e ~\bar \ell \gamma^\mu \ell ~ A_\mu \nn & & \\
{\cal L}_{CC} &=& \frac{g }{\sqrt{2}}  ~\bar \ell_L \gamma^\mu  \nu_L ~W_\mu^-+ h.c.   \no
\eea
where $\ell=\ell_L+\ell_R$ and $\theta_W$ is the Weinberg angle.  
 Experimentally it has been observed that at least two of the SM neutrinos have a small mass, 
not larger than the eV-scale \cite{Amsler:2008zzb}. 
In the following, we will account for the light neutrino masses and mixings by means of an effective Majorana
mass term, namely we add to the SM Lagrangian a dimension-5 non-renormalizable operator. 
Such a minimal extension of the SM is often referred to as the $3\nu$-SM.

Our aim here is to study the phenomenology of an additional heavy lepton family, with masses about the TeV-scale.
Thus, we add to the $3\nu$-SM matter content a 4th-family of leptons - for which we introduce the $\zeta$-flavor - 
composed by a lepton doublet, a charged lepton singlet and a gauge singlet:
\beq
L_\zeta= ({\nu_\zeta}_L ~~ \zeta_L)^T \sim (1,2,-1/2)~~~~,~~~~~{\zeta}_R\sim (1,1,-1)~~~,
~~~~~{\nu_\zeta}_R\sim (1,1,0)~~. \label{newLeptons}
\eeq
The $\zeta$-charged lepton, $\zeta = {\zeta}_L+{\zeta}_R$, will have a Dirac mass term like the other 
three charged leptons of the SM, but large enough to avoid conflict with the experimental limits 
(more on this later).  We work in the basis 
in which the $4\times4$ charged lepton mass matrix is diagonal.

\subsection{Heavy leptons with an exact flavor symmetry}

By imposing an exactly conserved new $\zeta$-lepton number we forbid, in this section, the mixing between the $\zeta$-neutrino and the three light neutrinos of the SM. 
The Lagrangian can be split as ${\cal L} = {\cal L}_{SM} + {\cal L}_\zeta$.

The Lagrangian mass terms we take for the $\zeta$-sector reads:
\begin{eqnarray}
{\cal L}_\zeta^{\rm mass} 
= -m_{\zeta} ~ \overline{ {\zeta}}~ {\zeta}  
- \frac{1}{2} \left[ \left( \begin{array}{cc}  \overline{\nu_{\zeta L}} & \overline{(\nu_{\zeta R})^c}  \end{array} \right)
\left( \begin{array}{cc}   0 & m_D \cr  m_D &  m_R \end{array} \right)  
\left( \begin{array}{c} (\nu_{\zeta L})^c \cr \nu_{\zeta R}  \end{array} \right) + h.c.\right] \ ,
\end{eqnarray}
Diagonalizing the neutrino mass matrix above, we obtain two independent Majorana eigenstates, $N_1$ and $N_2$, 
with real and positive masses, $M_1$ and $M_2$ (for convention $M_1 \le M_2$),
\beq
M_1=\frac{m_R}{2} \left( \sqrt{1+4\frac{m_D^2}{m_R^2}} -1 \right)~~~~,~~~~
M_2=\frac{m_R}{2} \left( \sqrt{1+4\frac{m_D^2}{m_R^2}} +1 \right)~~,
\eeq
which are related to the original Dirac and Majorana masses according to:
\begin{eqnarray}
M_1 M_2 = m^2_D \ , \qquad M_2-M_1 = m_R \ .
\end{eqnarray} 

The $\zeta$-neutrino chiral states will be an admixture of the two Majorana eigenstates $N_1$ and $N_2$:
\begin{equation}
\left( \begin{array}{c}
  \nu_{\zeta L} \cr  (\nu_{\zeta R})^c  \end{array} \right) = 
 \left( \begin{array}{cc}  i \cos\theta  & \sin\theta  \cr  -i \sin\theta & \cos\theta \end{array} \right) 
\left( \begin{array}{c}  P_L N_1 \cr P_L N_2  \end{array} \right) ~~~~,~~~~\tan2\theta = \frac{2m_D}{m_R} ~~.
\end{equation}
In the limit $m_D \ll m_R$ the seesaw mechanism would be at work (leading to $M_1\sim m_D^2 /m_R$, $M_2\sim m_R$, 
${\nu_{\zeta}}_L \sim  i  ~P_L N_1$, $({\nu_{\zeta R}})^c \sim  P_L N_2$). Here however we are more interested 
in the situation $m_D \sim m_R$, in which both Majorana neutrinos have a mass below the TeV-scale and hence have
a large $SU(2)$-active component. 

The neutral current interaction of the $\zeta$-leptons in terms 
of the heavy neutrino Majorana mass eigenstates reads:
\beq
{\cal L}^{NC}_\zeta = \frac{g }{\cos\theta_W}~ 
\left(\frac{1}{2} \overline{ {\nu_\zeta}_L} \gamma^\mu {\nu_\zeta}_L 
- \frac{1}{2} \overline{ \zeta_L} \gamma^\mu \zeta_L + \sin^2 \theta_W ~\bar \zeta \gamma^\mu \zeta \right) ~Z_\mu~ 
+ ~e  ~\bar \zeta \gamma^\mu \zeta ~A_\mu~~, 
\eeq
where
\beq
\label{neutral current}
\overline{ {\nu_\zeta}_L } \gamma^\mu {\nu_\zeta}_L= -\frac{\cos^2 \theta}{2} \bar N_1 \gamma^\mu \gamma_5 N_1 
-\frac{\sin^2 \theta}{2} \bar N_2 \gamma^\mu \gamma_5 N_2 + i \cos\theta \sin\theta  \bar N_2 \gamma^\mu  N_1 ~~.
\eeq
The interaction of the $Z$ with a couple of $N_{1}$ or $N_{2}$ is axial, while 
the one with two different $N_i$ is a vector interaction. 
As for the charged current:
\bea
{\cal L}^{CC}_\zeta 
    = \frac{g }{\sqrt{2}} W_\mu^- \bar \zeta_L \gamma^\mu (i \cos\theta ~P_L N_1 + \sin\theta ~P_L N_2)
    + h.c.~~.\no
\label{charged current no mix}
\eea 
The Dirac mass can be written in terms of the Yukawa coupling $y_{\zeta}$ and the Higgs vacuum expecation value $v$ as $ m_D= y_{\zeta}\,v/\sqrt{2}$. Hence the interaction of the 
new neutrinos with the Higgs field reads:
\bea
{\cal L}^{H}_\zeta 
&=& -\frac{ m_{D}}{v}  
\left( \overline{ {\nu_\zeta}_R}~ {\nu_\zeta}_L  + \overline{ {\nu_\zeta}_L}~ {\nu_\zeta}_R \right) ~H\\
&=& -\frac{ m_{D}}{v}  \left[  \cos\theta \sin\theta \left({\bar{N}_1}N_1 +{\bar{N}_2}N_2  \right) 
- i\,  \left( \cos^2\theta  - \sin^2\theta\right){\bar{N}_1}\gamma_5 N_2 \right] ~H~.  \no
\eea

\vskip .5cm

\subsection{Promiscuous heavy leptons}
\label{mixing}

\begin{figure}[h!]\begin{center}\begin{tabular}{c}
\includegraphics[width=8cm]{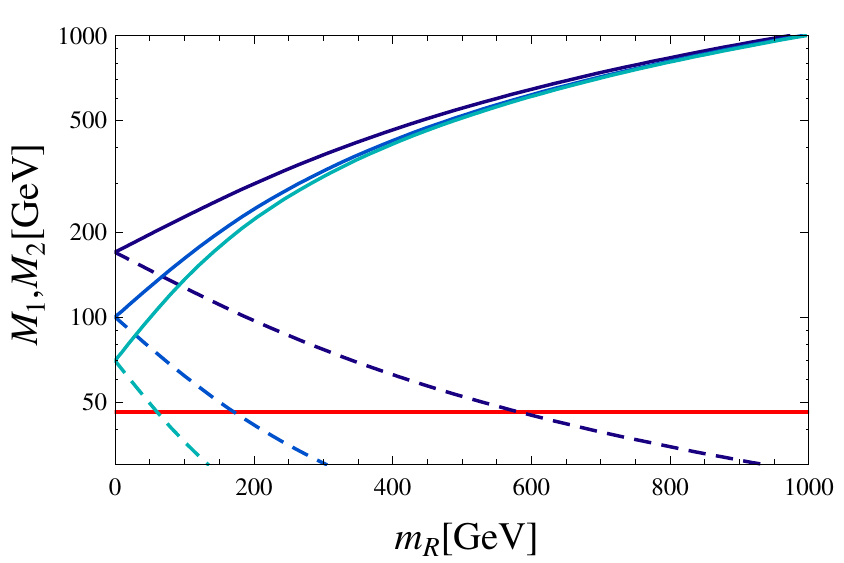}
\end{tabular}\end{center}\vspace*{-0.5cm} \caption{$M_1$ (dashed-lines) and $M_2$ (solid-lines)
as functions of $m_R$ for three representative values (from top to bottom) of $m'_D=\{170,100,70\}$ GeV. The red (solid) horizontal line marks the value $M_Z/2$.}
\label{fig-MM}
\end{figure}

In this section we consider the possibility that the new heavy leptons mix with the SM leptons.
{}For clarity of presentation we assume that the heavy neutrinos mix only with one SM neutrino of flavor $\ell$ ($\ell=e,\mu,\tau$) while we present the general case in appendix~\ref{sec:full mixing}. The entries of the mass matrix are:
\begin{equation}
-{\cal L}= \frac{1}{2} ( \begin{array}{ccc} 
 \overline{\nu_{\ell L}}& \overline{\nu_{\zeta L}} & \overline{(\nu_{\zeta R})^c}  \end{array} )
\left( \begin{array}{ccc}  {\cal O}(eV) & {\cal O}(eV) & m' \cr  {\cal O}(eV) &  {\cal O}(eV) & m_D \cr 
m' & m_D &  m_R \end{array} \right)  
\left( \begin{array}{c} (\nu_{\ell L})^c \cr (\nu_{\zeta L})^c \cr \nu_{\zeta R}  \end{array} \right) + h.c.~~.
\label{generalmatrix}
\end{equation} 
The measured values of the light 
neutrino masses suggest the entries of the upper 2$\times$2 block  to be of  ${\cal O}(eV)$ while the remaining entries are expected to be at least of the order of the electroweak energy scale. Given such a hierarchical structure and up to small corrections of ${\cal O}(eV/M_{1,2}) \lesssim 10^{-11}$,
one obtains the following form for the unitary matrix which diagonalises  eq.(\ref{generalmatrix}):
\begin{equation}
\left( \begin{array}{c}
\nu_{\ell L} \cr  \nu_{\zeta L} \cr  (\nu_{\zeta R})^c  \end{array} \right) =  V
\left( \begin{array}{c} P_L N_0 \cr P_L N_1 \cr P_L N_2  \end{array} \right) ~~,~~
V=\left( \begin{array}{ccc} \cos \theta' & i \cos\theta \sin\theta' & \sin\theta \sin\theta' 
\cr -\sin\theta' & i \cos\theta \cos\theta' & \sin\theta \cos\theta' \cr 
0 & -i \sin\theta & \cos\theta \end{array} \right).
\label{Eq: mixing matrix}
\end{equation}
$N_{0,1,2}$ are the new (Majorana) mass eigenstates and
\begin{equation}
\tan \theta' = \frac{m'}{m_D} ~~~, ~~~~~\tan 2 \theta= 2 ~\frac{m'_D}{m_R} ~~~,
~~~~~{m'_D}^2\equiv m_D^2+m'^2~~~.
\label{eqtheta}
\end{equation}
The light neutrino $N_0$ has a mass of ${\cal O}(eV)$. Up to corrections of ${\cal O}(eV)$, 
the heavy neutrinos $N_{1,2}$ have masses given by: 
\beq
M_1=\frac{m_R}{2} \left( \sqrt{1+4 \frac{{m'_D}^2}{m_R^2}} -1 \right) ~~,~~~~
M_2=\frac{m_R}{2} \left( \sqrt{1+4 \frac{{m'_D}^2}{m_R^2}} +1 \right) ~.
\label{eq:M1M2}
\eeq
In fig.~\ref{fig-MM} we display $M_1$ and $M_2$ as a function of $m_R$ for three representative 
values of $m'_D$ - from top to bottom $m'_D=\{170,100,70\}$ GeV. 
The smaller $m_R$ is the more the neutrinos $N_1$ and $N_2$ become (the two Weyl components of) a Dirac state. 

\noindent
Including the neutrino of flavor $\ell$, the fermion current in Eq.~(\ref{neutral current}) is replaced with 
\begin{eqnarray}
\bar {\nu_\ell}_L \gamma^\mu {\nu_\ell}_L+
\bar {\nu_\zeta}_L \gamma^\mu {\nu_\zeta}_L \nonumber &=& -\frac{1}{2} \bar N_0 \gamma^\mu \gamma_5 N_0 
-\frac{\cos^2 \theta}{2} \bar N_1 \gamma^\mu \gamma_5 N_1 
-\frac{\sin^2 \theta}{2} \bar N_2 \gamma^\mu \gamma_5 N_2 
\\
&+& i \cos\theta \sin\theta  \bar N_2 \gamma^\mu  N_1 \ ,
\label{eq:NCfull}
\end{eqnarray}
while the charged current terms in the Lagrangian, Eq.~(\ref{charged current no mix}), become 
\beq
{\cal L}^{CC}_\zeta = \frac{g }{\sqrt{2}} W_\mu^- \bar \zeta_L \gamma^\mu 
(-\sin\theta' P_L N_0 + i \cos\theta \cos\theta' ~P_L N_1 + \sin\theta \cos\theta' ~P_L N_2)+ h.c.
\label{eq:CCfull}
\eeq 
and
\beq
{\cal L}^{CC}_\ell = \frac{g }{\sqrt{2}} W_\mu^- \bar \ell_L \gamma^\mu 
(\cos\theta' P_L N_0 + i \cos\theta \sin\theta' ~P_L N_1 + \sin\theta \sin\theta' ~P_L N_2)+ h.c.
\label{eq:NCfull}
\eeq
Notice that the neutral current remains flavor diagonal at tree-level \footnote{The neutral current is not flavor diagonal in models with TeV scale right handed neutrinos involved in the see-saw mechanism for the light SM neutrino masses, see e.g. \cite{del Aguila:2005pf}}, hence the heavy neutrinos couple to the SM ones only through the charged current interactions at this order.
This is a distinctive feature of our TeV neutrino physics. The SM like Yukawa interactions lead to the following terms involving the Higgs: 
\bea
&-&\frac{m_{D}}{v} 
 \left( \overline{ {\nu_\zeta}_R}~ {\nu_\zeta}_L  + \overline{ {\nu_\zeta}_L}~ {\nu_\zeta}_R \right) H
=  - \frac{m_{D}}{v} 
[   \sin\theta \cos\theta \cos\theta' ({\bar{N}_1}N_1 +{\bar{N}_2}N_2) \\
& ~&~~~ - i \cos\theta'( \cos^2\theta  - \sin^2\theta ) {\bar{N}_1}\gamma_5 N_2 
- i \sin\theta \sin\theta' \bar{N}_1 \gamma^5 N_0 -  \cos\theta \sin\theta' \bar{N}_2 N_0 ]~ H~ \no
\label{Eq:Yukawa1's}
\eea
and
\bea
&-&\frac{m'}{v} 
 \left( \overline{ {\nu_\zeta}_R}~ {\nu_\ell}_L  + \overline{ {\nu_\ell}_L}~ {\nu_\zeta}_R \right) H
=  - \frac{m'}{v} 
[   \sin\theta \cos\theta \sin\theta' ({\bar{N}_1}N_1 +{\bar{N}_2}N_2) \\
& ~&~~~ - i \sin\theta'( \cos^2\theta  - \sin^2\theta ) {\bar{N}_1}\gamma_5 N_2 
+ i \sin\theta \cos\theta' \bar{N}_1 \gamma^5 N_0 +  \cos\theta \cos\theta' \bar{N}_2 N_0 ]~ H~ \no~.
\label{Eq:Yukawa2's}
\eea
The sum of these two expressions is equal to:
\bea
  - \frac{m_D}{v}  (1+\frac{m'^2}{m_D^2}) \cos\theta'
[   \sin\theta \cos\theta  ({\bar{N}_1}N_1 +{\bar{N}_2}N_2) 
  - i ( \cos^2\theta  - \sin^2\theta ) {\bar{N}_1}\gamma_5 N_2 ]~ H~ .
\label{Eq:Yukawa's}
\eea

\subsection{Heavy Leptons in Minimal Walking Technicolor models}
The simple model presented above is per se inconsistent because of uncanceled gauge and Witten anomalies  \cite{Witten:1982fp}. To avoid such an inconsistency we add new fermions charged under the electroweak gauge group whose additional new gauge dynamics is responsible for a dynamical breaking of the electroweak symmetry. The model we use as a template is Minimal Walking Technicolor (MWT) \cite{Sannino:2004qp}.

MWT is an SU(2) technicolor gauge theory 
with two adjoint
technifermion:
 \beq Q_L^a=\left(\begin{array}{c} U^{a} \\D^{a} \end{array}\right)_L , \qquad U_R^a \
, \quad D_R^a \ ,  \qquad a=1,2,3 \ ,\eeq with $a$ being the adjoint color index of SU(2). The left handed fields 
are arranged in three
doublets of the SU(2)$_L$ weak interactions in the standard fashion. The condensate is 
$\langle \bar{U}U + \bar{D}D \rangle$ which
correctly breaks the electroweak symmetry.
Anomalies are canceled by adding a heavy Lepton doublet \cite{Dietrich:2005jn}: 
\beq L_L =
\left(
\begin{array}{c} N \\ E \end{array} \right)_L , \qquad N_R \ ,~E_R \
. \eeq In general, the gauge anomalies cancel using the following
generic hypercharge assignment
\begin{align}
Y(Q_L)=&\frac{y}{2} \ ,&\qquad Y(U_R,D_R)&=\left(\frac{y+1}{2},\frac{y-1}{2}\right) \ , \label{assign1} \\
Y(L_L)=& -3\frac{y}{2} \ ,&\qquad
Y(N_R,E_R)&=\left(\frac{-3y+1}{2},\frac{-3y-1}{2}\right) \ \label{assign2} ,
\end{align}
where the parameter $y$ can take any real value \cite{Dietrich:2005jn}. One recovers the SM-like hypercharge
assignment for $y=1/3$. If we choose indeed the SM hypercharge assignment the new lepton family can be identified with the heavy family discussed above with $N=\nu_{\zeta}$ and $E=\zeta$.  The physics of the strongly coupled sector of MWT has been investigated already in the literature \cite{Sannino:2004qp,Dietrich:2005jn,Foadi:2007ue,Dietrich:2009ix} while the focus here is on the heavy leptons, their interplay with the new strong dynamics as well as the mixing with the SM light neutrinos. {We note that in addition to the appearance of fundamental heavy leptons canceling the gauge anomalies in the MWT model, both the MWT and the Ultra Minimal Walking Technicolor model feature composite heavy leptons \cite{Foadi:2007ue,Ryttov:2008xe}}      

\noindent
The low energy effective theory we will use for determining the interesting signals for LHC phenomenology \cite{Foadi:2007ue,Appelquist:1999dq} contains composite spin one and spin zero states and we summarize it in the appendix. The new heavy spin one states will mix with the SM gauge bosons and hence modify the charged and flavor currents in  Eqs.~(\ref{eq:CCfull}) and Eqs.~(\ref{eq:NCfull}).

%%%%%%%%%%%%%%%%%%%%%%%%%%%%%%%%%%%%%%%%%%%%%%%%%%%%%%%%%%%%%%%%%%%%%%%%%%%%%%%%%%%%%%%%%%%%%%
\section{Parameter space constraints}

There are both direct and indirect constraints from collider experiments as well as cosmological constraints on heavy leptons, 
which we discuss below.

%%%%%%%%%%%%%%%%%%%%%%%%%%%%%%%%%%%%%%%%%%%%
\subsection{Limits from LEP and Tevatron}

\vskip .4cm
\underline{LEP}

\noindent
For a neutral fourth lepton decaying to SM leptons, mass limits at 95 $\%$ CL from LEP2 were determined in \cite{Achard:2001qw} including 
data up to $\sqrt{s} = 208$ GeV. 
The study assumed decays of the heavy neutrino into $\ell W$ and assumed that the relevant mixing coefficients square are larger than $10^{-11}$
(such that the heavy neutrino decays inside the detector). The best mass limit was achieved for the $W \mu$ decay mode yielding a mass limit of $101.5$ GeV for a Dirac neutrino and $90.7$ GeV 
for a Majorana neutrino, while the $ W \tau$ decay mode yields the weakest mass limit of $90.3$ GeV for a Dirac neutrino and $80.5$ GeV for a Majorana neutrino. The study also assumed a pair production cross-section of the $N_1$ corresponding to $\cos \theta =1$ in Eq.~\ref{eq:NCfull} as also pointed out in \cite{Atre:2009rg}. Therefore the lower bound on $M_1$ is reduced if  $\cos \theta < 1$.
With no assumption on the size of mixing coefficients an earlier mass limit at 95 $\%$ CL from LEP1 for a decaying heavy neutrino was obtained in \cite{Abreu:1991pr} from the study of the total Z decay width yielding 44 GeV for a Dirac neutrino and 38.2 GeV for a Majorana neutrino.

In \cite{Achard:2001qw} mass limits at 95 $\%$ CL were also set on a charged fourth lepton.{} For a charged decaying fourth lepton, the mass limit achieved is $100.8$ GeV for decays into $\nu W$ and 
$101.9 $ GeV for decays into $N_1 W$ (assuming the mass splitting between $N_1$ and $\zeta$
is at least $15$ GeV). For a stable charged lepton the mass limit achieved is $102.6$ GeV. 

For a stable heavy neutrino mass limits arise from the study of the invisible Z decay width presented below.

\vskip .4cm
\underline{Tevatron}

\noindent
At Tevatron, the CDF collaboration has also searched for single, weakly interacting and long-lived charged particles, or more generally CHAMPs 
(CHArged Massive Particles). They find an upper limit to the production cross section 
(times acceptance) of  $10$ fb at $\sqrt{s}=1.96$ TeV at 95 $\%$CL \cite{Aaltonen:2009ke}. For our CHAMP candidate $\zeta^\pm$ we are below this bound for masses above $\sim 100$ GeV. 

\vskip.4cm
\underline{Invisible width of the $Z$}

\begin{figure}[h!]\begin{center}\begin{tabular}{c}
\includegraphics[width=8.15cm]{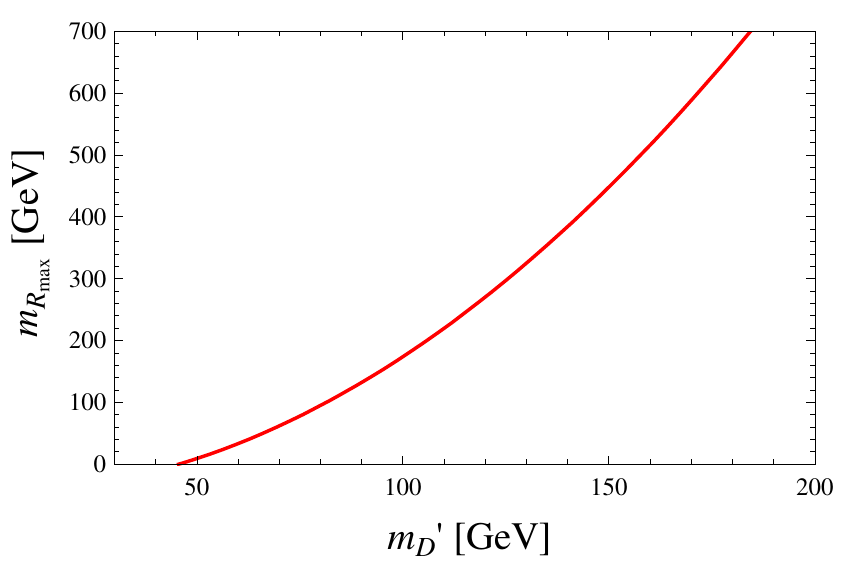}~~~~~
\includegraphics[width=8cm]{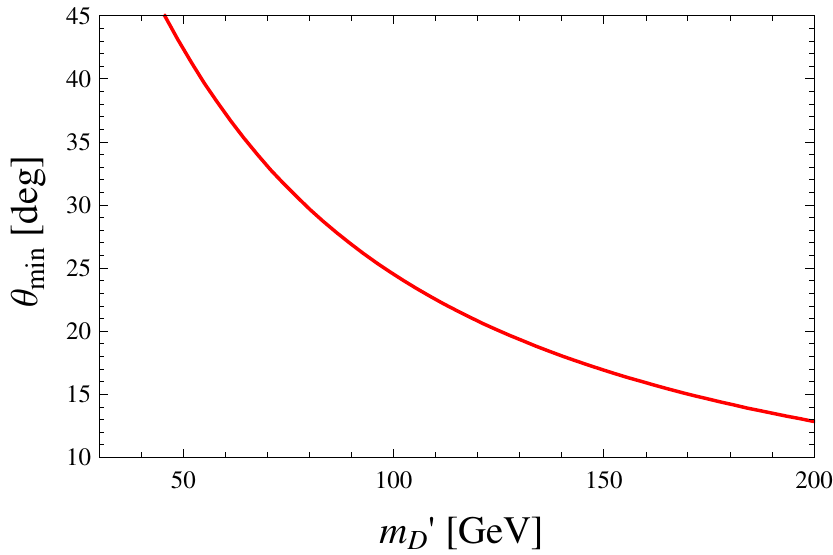}
\end{tabular}\end{center}\vspace*{-0.5cm} \caption{Left: 
Upper bound on $m_R$ as a function of $m'_D$, eq. (\ref{mrub}).
Right: Lower bound on $\theta$ as a function of $m'_D$, eq. (\ref{tmin}).}
\label{fig-bounds}
\end{figure}

\noindent
From eq. (\ref{eq:NCfull}), it follows that $M_1 > M_Z/2$ is a sufficient requirement 
to forbid any contribution from the heavy neutrinos $N_{1,2}$ to the $Z$ decay width.   
As can be seen from eq.(\ref{eq:M1M2}), such a requirement implies an upper bound on $m_R$: 
\beq
m_R \le {m_R}_{max}= 2~\frac{{m'_D}^2}{m_Z} - \frac{m_Z}{2}~~. 
\label{mrub}
\eeq 
As can be also seen from fig.\ref{fig-MM}, for $m_R=0$ this implies $m'_D \ge M_Z/2$. 
From eqs.(\ref{eqtheta}) and (\ref{mrub}), it turns out that the range of 
values allowed for $\theta$ is:
\beq
 \theta_{min}\equiv \arctan \frac{m_Z}{2 m'_D} \le \theta  \le 45^\circ~~.
 \label{tmin}
\eeq 
The upper bound on $m_R$ and the lower bound on $\theta$ are shown respectively in the left and right 
plots of fig. \ref{fig-bounds}, as a function of $m'_D$. 
   
\begin{figure}[h!]
\begin{center}\begin{tabular}{c}
\includegraphics[width=7cm]{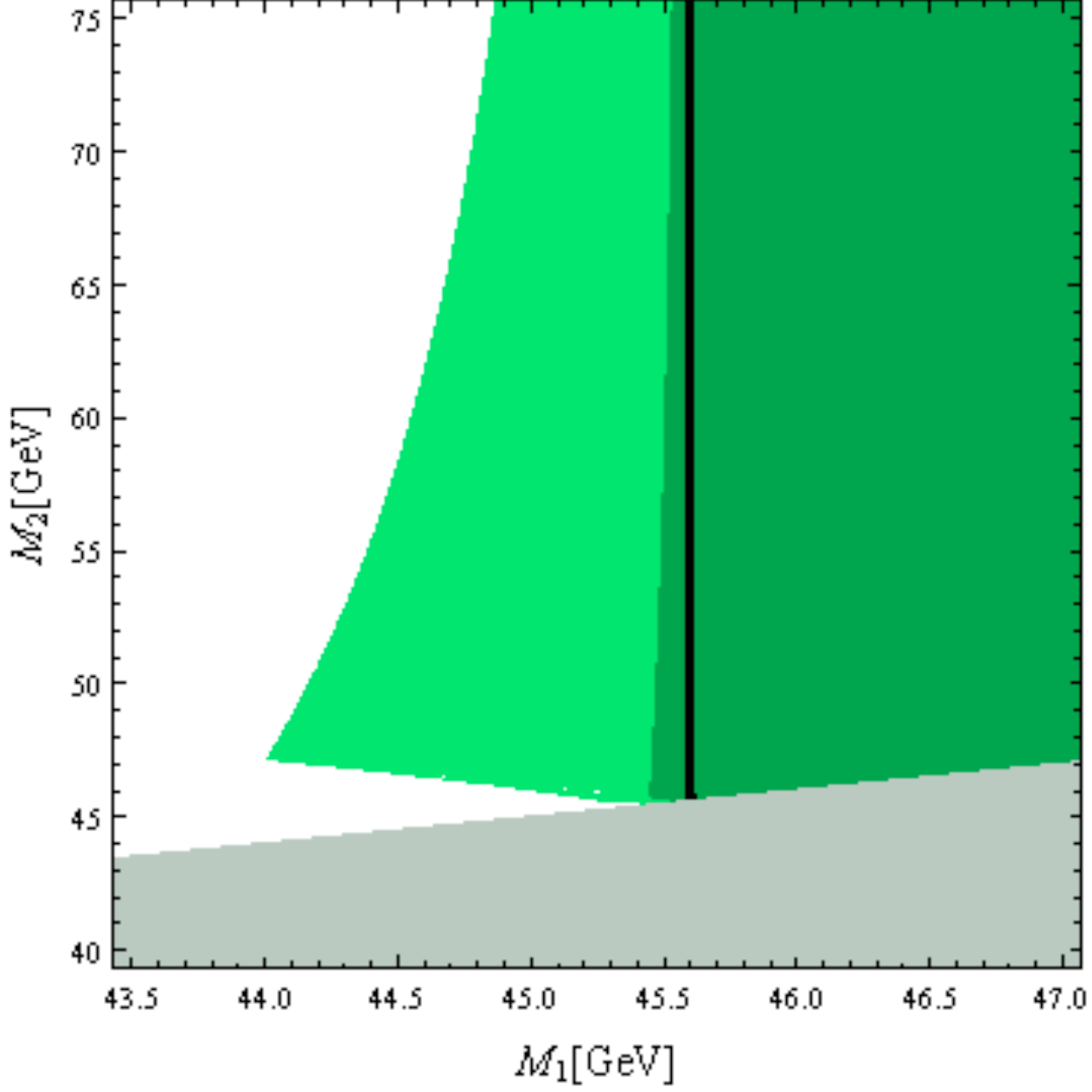} 
\end{tabular}\end{center}\vspace*{-0.5cm} 
\caption{Allowed region at 3 (light green) and 2 $\sigma$ (dark green) in the $(M_1,M_2)$ plane. 
The shaded gray region is excluded by the condition $M_2>M_1$. The black line corresponds to 
$\langle N \rangle=3$, namely $M_1 \leq M_Z/2$. 
 }\label{fig-Z}
\end{figure}

One could however wonder whether the scenario $M_1 < M_Z/2 < M_2$ is still viable, because of 
the  $\cos^2\theta$ suppression in eq.(\ref{eq:NCfull}). 
Indeed, by adding the pair of Majorana neutrinos $N_1$ and $N_2$ to the three light neutrinos 
of the SM, the effective number of neutrinos involved in the $Z$ invisible width, 
$\langle N \rangle= \Gamma_{inv}/\Gamma_0$ (where $\Gamma_0=\Gamma(Z\rightarrow \bar \nu \nu)$ is 
the amplitude for $Z$ to decay into a light, practically massless, neutrino), becomes \cite{Jarlskog}:  
\bea
\langle N \rangle &=& 3~ + ~\theta(M_Z-2 M_1) ~\cos^4\theta~ x_{11} 
+ ~\theta(M_Z-M_1-M_2)~2\cos^2\theta \sin^2\theta~ x_{12} \nn 
& ~&~~~~~~~~~~~~~~~~~~~~~~~~~~~+ \theta(M_Z-2 M_2) ~\sin^4\theta ~x_{22}~~~~~,
\eea
where $\theta(x)$ is the step function and
\beq
x_{ij} =\sqrt{1+ \left(\frac{M_i}{M_Z}\right)^4 + \left(\frac{M_j}{M_Z}\right)^4 
- 2 \left[ \left(\frac{M_i}{M_Z}\right)^2 +\left(\frac{M_j}{M_Z}\right)^2 + \frac{M_i^2 M_j^2}{M_Z^4}\right]}~~
~~~~,~~i,j=1,2~~.
\eeq
From the direct measurement of the invisible $Z$ width, we have at present 
$\langle N \rangle =2.92\pm 0.05$ \cite{Amsler:2008zzb}. 
We display in fig. \ref{fig-Z} the regions of the $(M_1,M_2)$ plane allowed at 3 and 2 $\sigma$ 
(light and dark green respectively).
Notice that, at 3 $\sigma$, $M_1$ is allowed to be smaller than $M_Z/2$ by less than a couple
of GeV. 

Actually, even when the $Z$ decays occur below the mass threshold for the production of the heavy states $N_{1,2}$,
there could be a deviation from the SM prediction $\langle N \rangle=3$. This is because  
the standard gauge eigentate $\nu_{\ell L }$ of eq. (\ref{Eq: mixing matrix}) is effectively replaced 
by its normalised projection into the subspace of the light neutrinos, $P_L N_0$. Since the SM neutrino
of flavor $\ell$ mixes with the heavy neutrino $\nu_{\zeta L}$ belonging to a lepton doublet,  
the effective number of neutrinos turns out to be slightly increased \cite{Langacker:1988ur, Nardi:1991rg}.
At leading order in $\theta'$ one has:
\beq
\langle N \rangle = 3 + \frac{3}{2} \sin^2\theta'~~~,
\eeq 
which applies when the mixing involves $\ell=e,\mu$ (there would be no correction for $\ell=\tau$). 
Given that \cite{Amsler:2008zzb} $\langle N \rangle =2.92\pm 0.05$, one learns that a possible mixing with the 
neutrinos of flavor $e$ and $\mu$ is subject to the bound $\sin^2\theta' \le 0.047$ at 3 $\sigma$. 
Such a bound is however weaker than the one following from lepton universality, that we now discuss.

%%%%%%%%%%%%%%%%%%%%%%%%%%%%%%%%%%%%%%%%%%%%%%%%%%%%%%%%%%%%%%%%
\subsection{Direct constraints on mixing}

The mixing angle $\theta'$ of the heavy leptons with one SM lepton family (we consider each family in turn) 
is strongly constrained by lepton universality tests \cite{Langacker:1988ur, Nardi:1991rg}. {}In addition for the second family there are further constriaints when imposing unitarity of the CKM matrix.

The current status of lepton universality tests is summarised in \cite{Pich:2008ni}.
The present data verify the universality of the leptonic charged-current couplings to the 0.2\% level.
For the mixings with the neutrinos of flavor $e$ and $\mu$, one can extract an upper limit 
on $\sin^2\theta'$ of about $0.012$ at 3 $\sigma$, while for the mixing with the $\tau$ of about 
$0.15$ at 3 $\sigma$. 

The CKM unitarity constraint applies only to the mixing with a neutrino of flavor $\mu$ and 
reads $\sum_i |V_{ui}|^2 = 1/\cos^2\theta'$. Given the present determination \cite{Amsler:2008zzb}
$\sum_i |V_{ui}|^2 = 0.9999 \pm 0.0011$, one derives that $\sin^2\theta' \le 0.0033$ at about 3 $\sigma$.
It should be stressed that the CKM unitarity constraint above can be modified if the charged leptons 
or the quarks mix with additional exotic particles.

%%%%%%%%%%%%%%%%%%%%%%%%%%%%%%%%%%%%%%
\subsection{Indirect constraints from EW precision data}

Additional constraints come from Electroweak precision data. The S and T parameters \cite{Peskin:1991sw} 
for a sequential fourth lepton doublet and a right-handed singlet were computed in \cite{Gates:1991uu} 
including both Dirac masses for the doublet and Majorana masses for the right-handed singlet, as in our setup. 
We give the formulas for arbitrary hypercharge in Appendix \ref{gates}. 

The heavy leptons are part of the MWT model,  we also consider the contribution 
from the technicolor sector. Using the one-loop perturbative inspired naive estimate of $S_{\rm naive}$ we have 
\bea
S_{\rm naive} = \frac{1}{6\pi} d(R) N_D \ ,
\eea
where $d(R)=3$ is the dimension of the techni-fermion representation $R$ for MWT while $N_D=1$ is the number of 
electroweak techni-doublets. In a walking Technicolor theory the non-perturbative contributions can futher reduce the $S$ parameter value \cite{Sundrum:1991rf,Appelquist:1998xf,Kurachi:2006mu}.
In this study we will take $S=0.1$ from the technicolor sector following \cite{Foadi:2007ue}. In addition to the $S$ and $T$ parameters coming from the new strong dynamics we also have the contribution from the heavy lepton sector. At one loop these two contributions are additive and we will consider from together in the following discussion.

\begin{figure}
{\includegraphics[height=6.5cm,width=7.82cm]{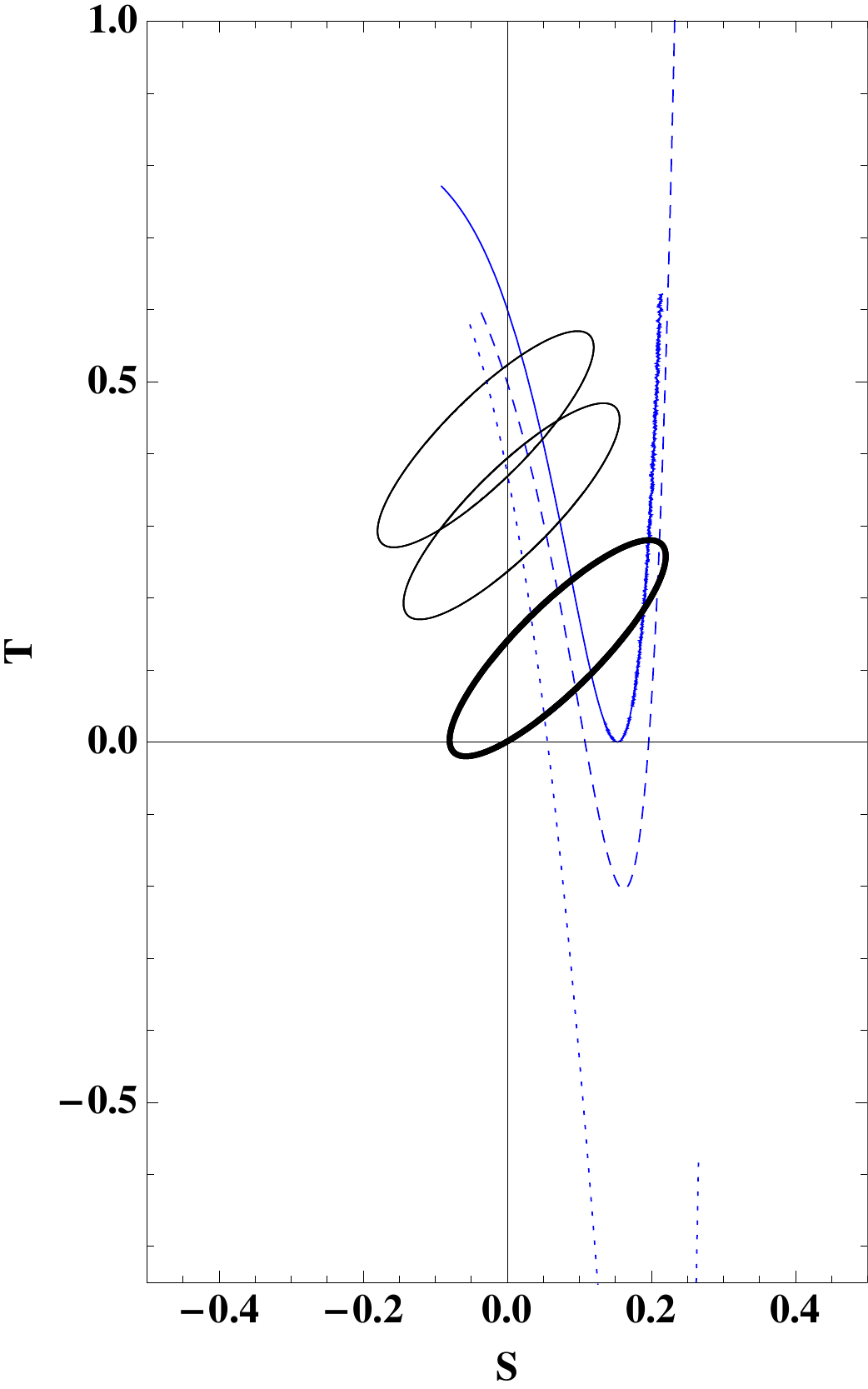}
\put(-150,200){\Large $M_\zeta=250$ GeV} \put(50,200){\Large $M_\zeta=500$ GeV}
\includegraphics[height=6.5cm,width=7.82cm]{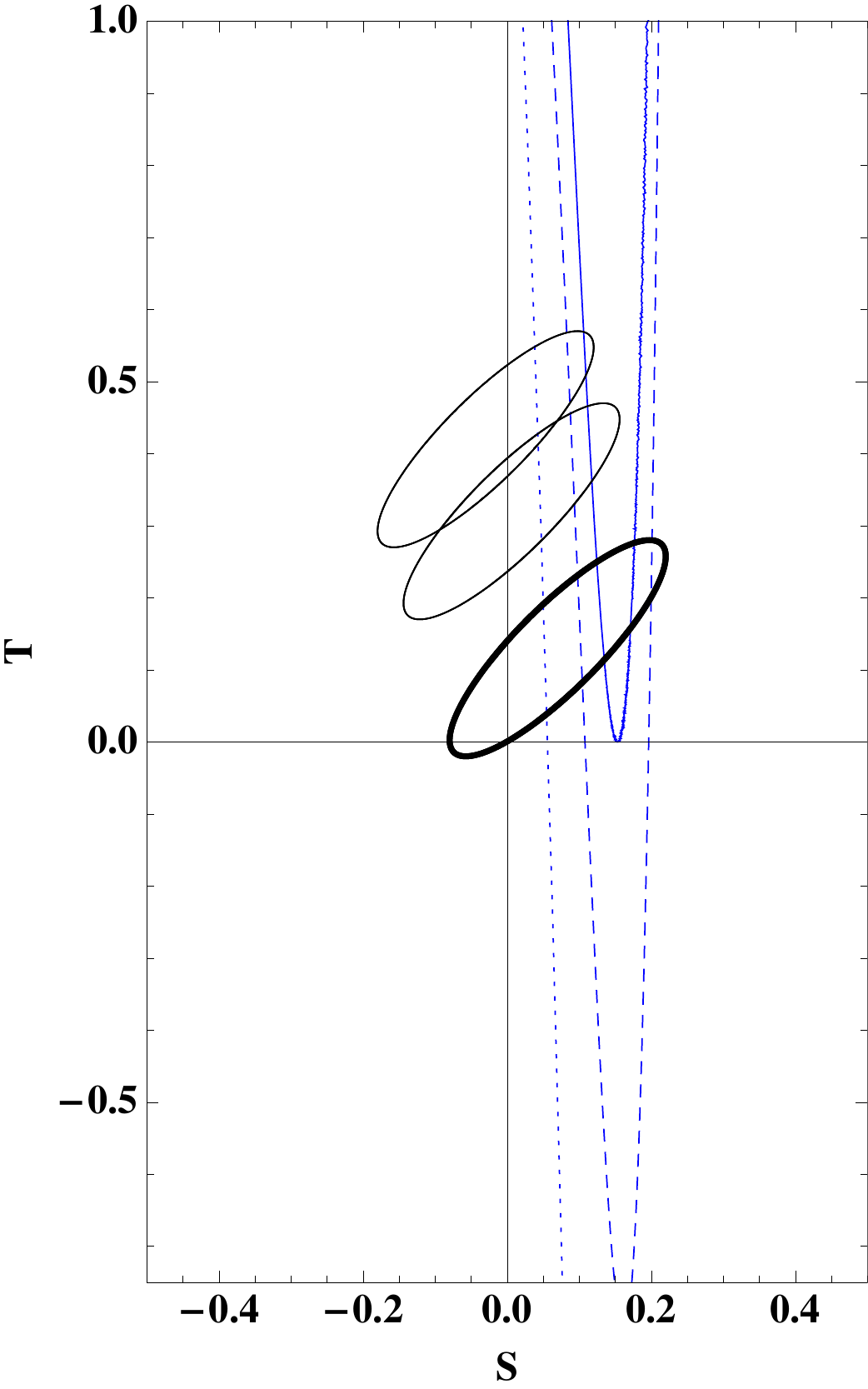}
}
\caption{S and T in the MWT model as a function of $M_1/M_\zeta$. 
We assume $S_{MWT}=\frac{1}{2\pi}, \ T_{MWT}=0$. $M_\zeta=250, 500$ GeV on the left, right plots respectively 
and $m_R/M_\zeta=0, 1, 5$ for solid, dashed, dotted lines respectively. $M_1/M_\zeta$ is varied from 0.1 
(on the upper left branch) to 1.75 (on the upper right branch) along the hyperbolae. 
The thick black ellipsis correspond to the 68 \% CL confidence level fit given in \cite{:2005ema} 
for $M_H=150$ GeV and $m_t=175$ GeV. The other two ellipses correspond to $M_H=0.5, 1$ TeV respectively. 
}
\label{fig:SandTellipse}
\end{figure}

In Fig.~\ref{fig:SandTellipse} we plot the S and T parameters of the MWT model against the experimentally allowed
values at the 68 \% CL from \cite{:2005ema}. The reference point at which S,T and U vanish 
was taken to be $M_H=150$ GeV and $m_t=175$ GeV. The lower ellipsis in  Fig.~\ref{fig:SandTellipse} reproduces the one in \cite{:2005ema}. 
The other two ellipses corresponds to simply propagating the first ellipsis to $M_H=0.5$ TeV and $M_H=1$ TeV, 
using the S and T contribution from the heavy Higgs as in \cite{Peskin:1991sw}. Note that redoing the full fit 
for the heavy Higgs reference masses results in error ellipses towards more negative S for $M_H=0.5,1$ TeV 
in \cite{Amsler:2008zzb}. Also note that the central value for $S$ in \cite{Amsler:2008zzb} is small 
and negative as opposed to the small and positive central value from the fit in \cite{:2005ema} we use here.
We tabulate the values of $\frac{M_1}{M_\zeta}$ for which S and T are within the 68 \% CL ellipses 
(from the upper edge of the top ellipsis to the lower edge of the bottom ellipsis) in table ~\ref{table:1}.

\begin{table}[htp]\vskip .5 cm
\begin{tabular}{c c||c|c|c}
& & $m_R=0$ & $m_R=M_\zeta$ & $m_R=5 M_\zeta$
\\ 
& & & & \\  \hline \hline  & & & & \\
&$M_\zeta=250$ GeV &0.28 - 0.72, 1.4 - 1.5 & 0.12 - 0.35, 1.3 - 1.33 & 0.12 - 0.25     \\
& & & & \\ \hline & & & &\\
&$M_\zeta=500$ GeV & 0.65 - 0.85, 1.2 - 1.25 & 0.25 - 0.4, 1.17 - 1.19 & 0.21 - 0.25   \\ & & & &
\end{tabular}
\caption{Table of allowed value of $\frac{M_1}{M_\zeta}$ within the 68 CL limit for 3 
values $m_R=M_2-M_1$ and 2 values of $M_\zeta$ with respect to the solid confidence ellipses, 
corresponding to $M_H=150$ GeV and $m_t=175$ GeV \cite{:2005ema} }.
\label{table:1}\vskip .5cm
\end{table}

In general the lepton masses are less constrained by the Electroweak precision measurements in the Dirac 
limit $M_2=M_1$ and the measurements do not rule out $M_\zeta < M_1$.

The EW constraints for the MWT model with only a Dirac mass for the 4th lepton were considered in detail in 
\cite{Dietrich:2005wk,Dietrich:2005jn,Foadi:2007ue} while the electroweak parameters in which a Majorana 
mass is given to the left-handed neutrino were discussed in \cite{Holdom:1996bn}.

%%%%%%%%%%%%%%%%%%%%%%%%%%%%%%%%%%%%%%%%%%
\subsection{Cosmological constraints}

We first consider $M_\zeta > M_1$. In this regime $\zeta^\pm$ decays immediately to $N_1 W$. 
Instead $N_1$ only decays through mixing with the SM leptons and could in principle be long lived or stable. 
Assuming the heavy neutrinos are produced as expected in the Big Bang model and that they clump as ordinary baryons, 
the lower mass limits from CDMS II are $500$ GeV for a Dirac neutrino and no exclusion above $\frac{1}{2} M_Z$ 
for a Majorana neutrino \cite{Rybka:2005vv}. The lower limit for the Dirac neutrino might be significantly 
reduced if the neutrinos do not clump like ordinary baryons.  
If $N_1$ is responsible for the dark matter density the mass limits change. 
This scenario was studied in \cite{Kainulainen:2006wq,Kouvaris:2007iq}. It is worth mentioning that MWT models can allow for different type of dark matter candidates \cite{Gudnason:2006ug,Gudnason:2006yj} of possible interest to recent cosmological observations \cite{Nardi:2008ix}.

Cosmological constraints for $M_\zeta < M_1$ were considered in \cite{Sher:1992yr}. In this regime 
$\zeta^\pm$ will only decay through mixing with the SM leptons and this means $\zeta^\pm$ could be long lived, if $\sin\theta'$ is sufficiently small in Eq.~(\ref{eq:CCfull}).
It is excluded that $\zeta^\pm$ could be absolutely stable since this would result in heavy hydrogen, 
which has not been observed.
For a $100-500$ GeV charged lepton, the upper limit on the life-time is constrained from big-bang nucleosynthesis 
\cite{Ellis:1984er,Ellis:1985fp,Ellis:1990nb,Holtmann:1998gd} to be below $10^7 - 10^8$ sec as discussed in \cite{Frampton:1999xi}.

\section{LHC phenomenology}
In this section we investigate aspects of the phenomenology related to the interplay between the new weekly coupled sector, i.e. the heavy leptons with its mixing with the SM fermions, and the new strongly coupled sector breaking the electroweak symmetry dynamically. We consider only the MWT global symmetries relevant for the electroweak sector, i.e. the subsector  $SU(2)\times SU(2)$ spontaneously breaking to $SU(2)$. The low energy spectrum contains, besides the composite Higgs, two $SU(2)$ triplets of (axial-) vector spin one mesons.  The effective Lagrangian has been introduced in \cite{Foadi:2007ue,Belyaev:2008yj} and summarized in the Appendix \ref{effective} along with the basic input parameters.  The spin one massive eigenstates are indicated with $R_{1}$ and $R_2$ and are linear combinations of the composite vector/axial mesons of MWT and the weak gauge boson eigenstates. 
{ We have implemented the $SU(2)\times SU(2)$ technicolor sector in CalcHEP \cite{Pukhov:2004ca}
using the LanHEP module \cite{Semenov:2008jy} to generate the Feynman rules in \cite{Belyaev:2008yj}. We have added the new leptons to this implementation for the present study.}

%%%%%%%%%%%%%%%%%%%%%%%%%%%%%%%%%%%%%%%%%%%%%%%%%%%%%%%%%%%%%%%%%%%%
\subsection{Production and decay of the new leptons }
\begin{figure}
{\includegraphics[height=3cm,width=6cm]{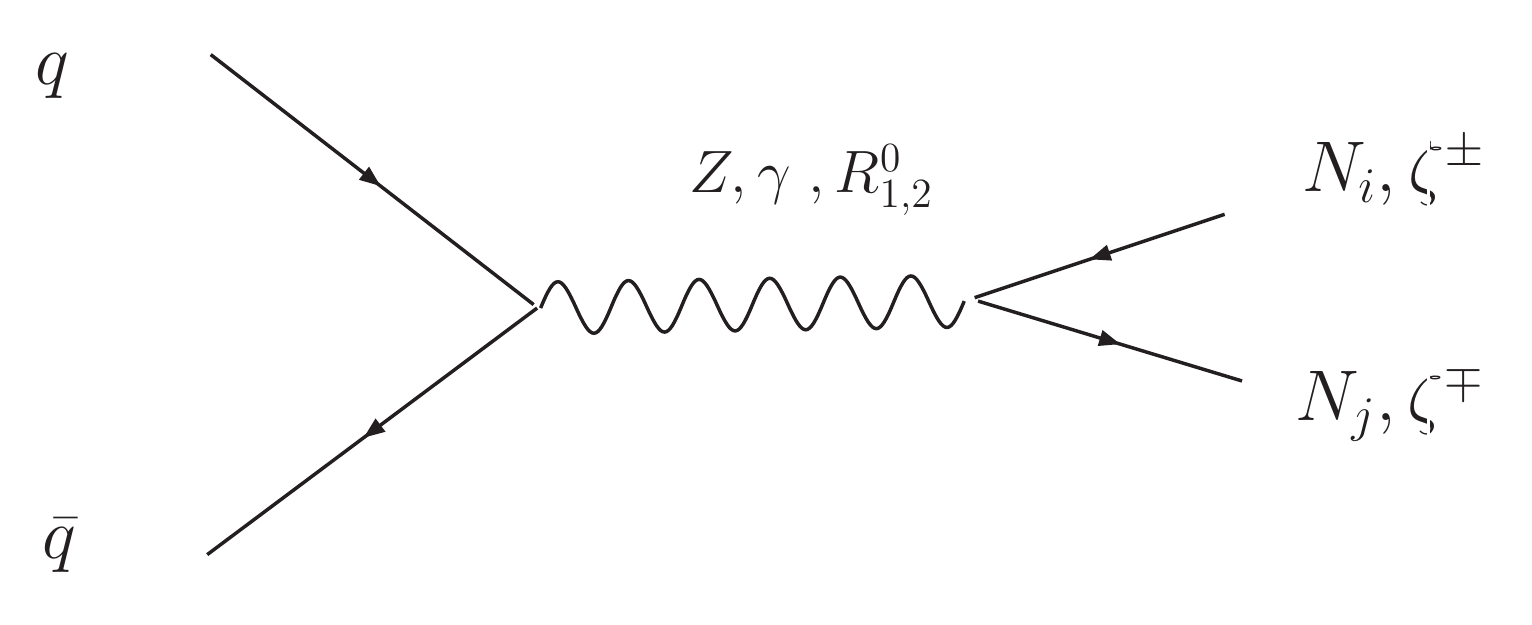}
\includegraphics[height=3cm,width=6cm]{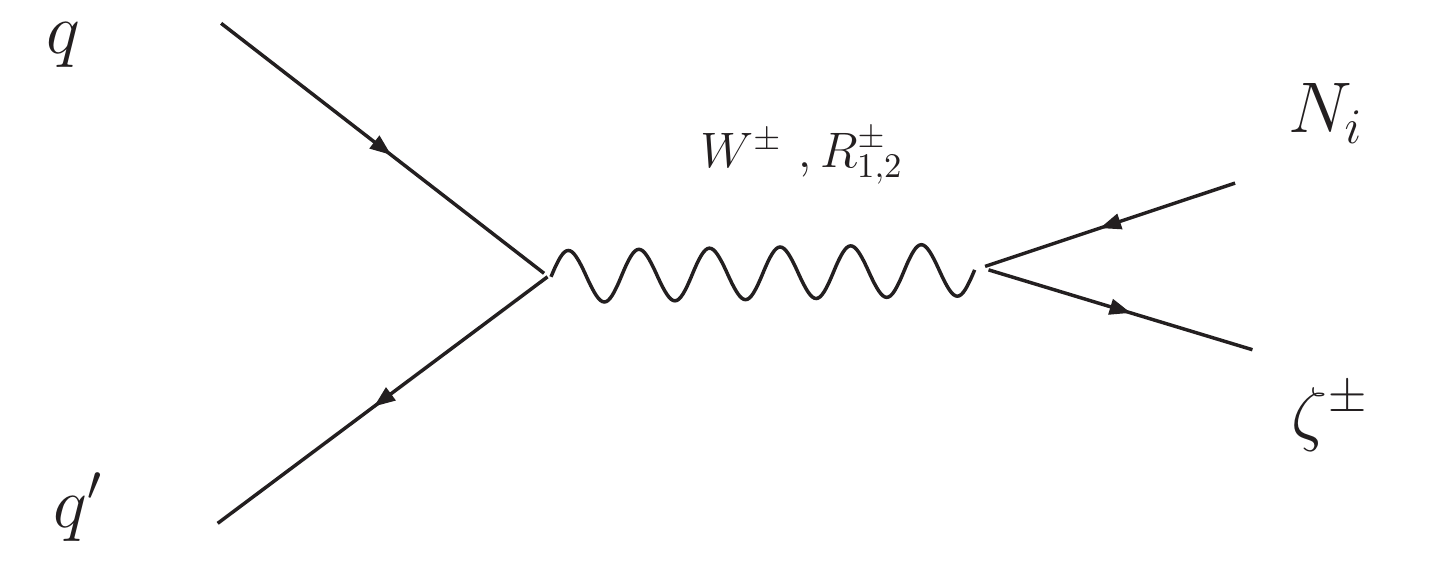}
}
\caption{Feynman diagrams for direct production of the heavy leptons in MWT with $i,j=1,2.$}
\label{fig:feynman diagrams}
\end{figure}

\begin{figure}
{\includegraphics[height=6.5cm,width=7.82cm]{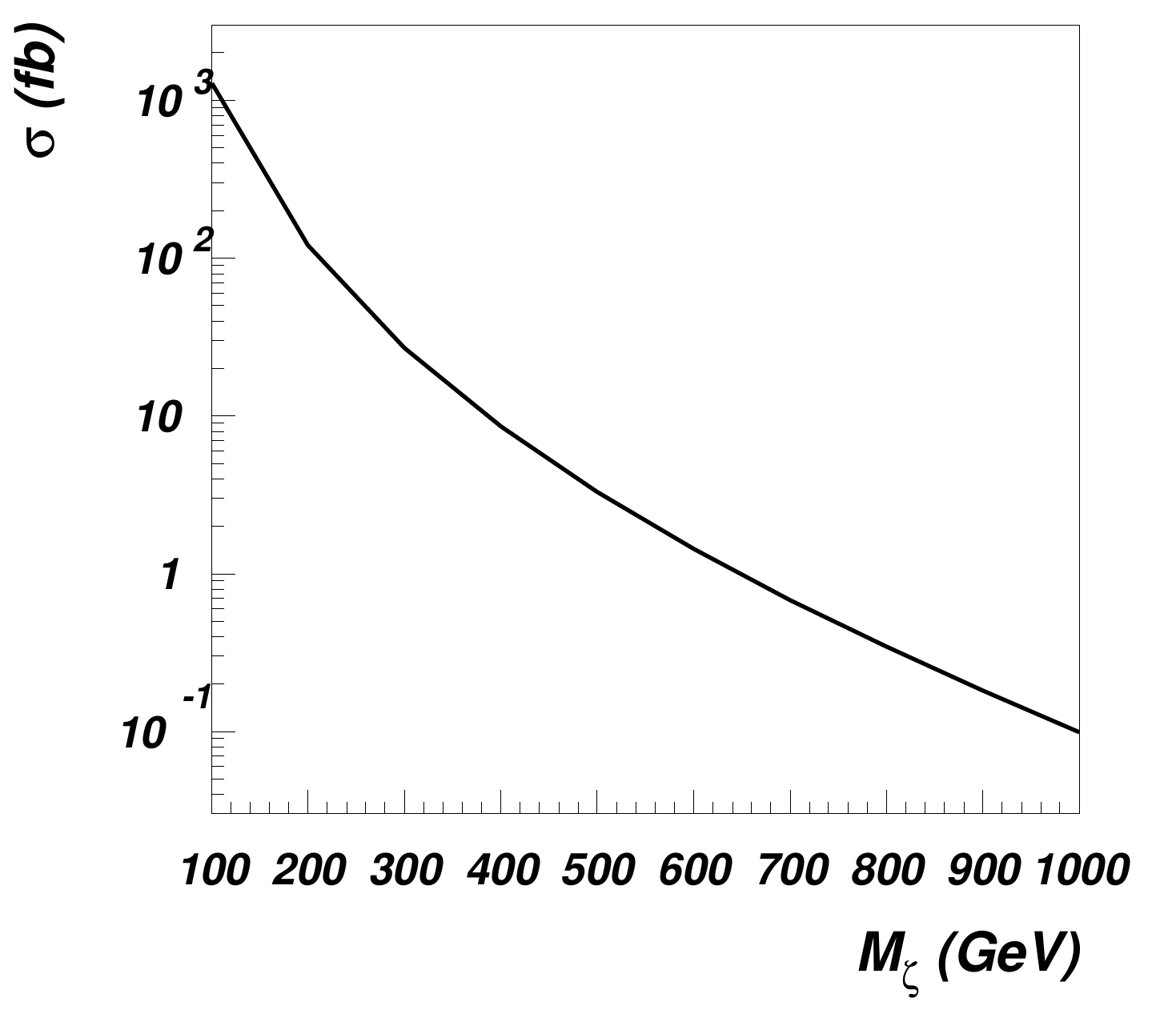}
\includegraphics[height=6.5cm,width=7.82cm]{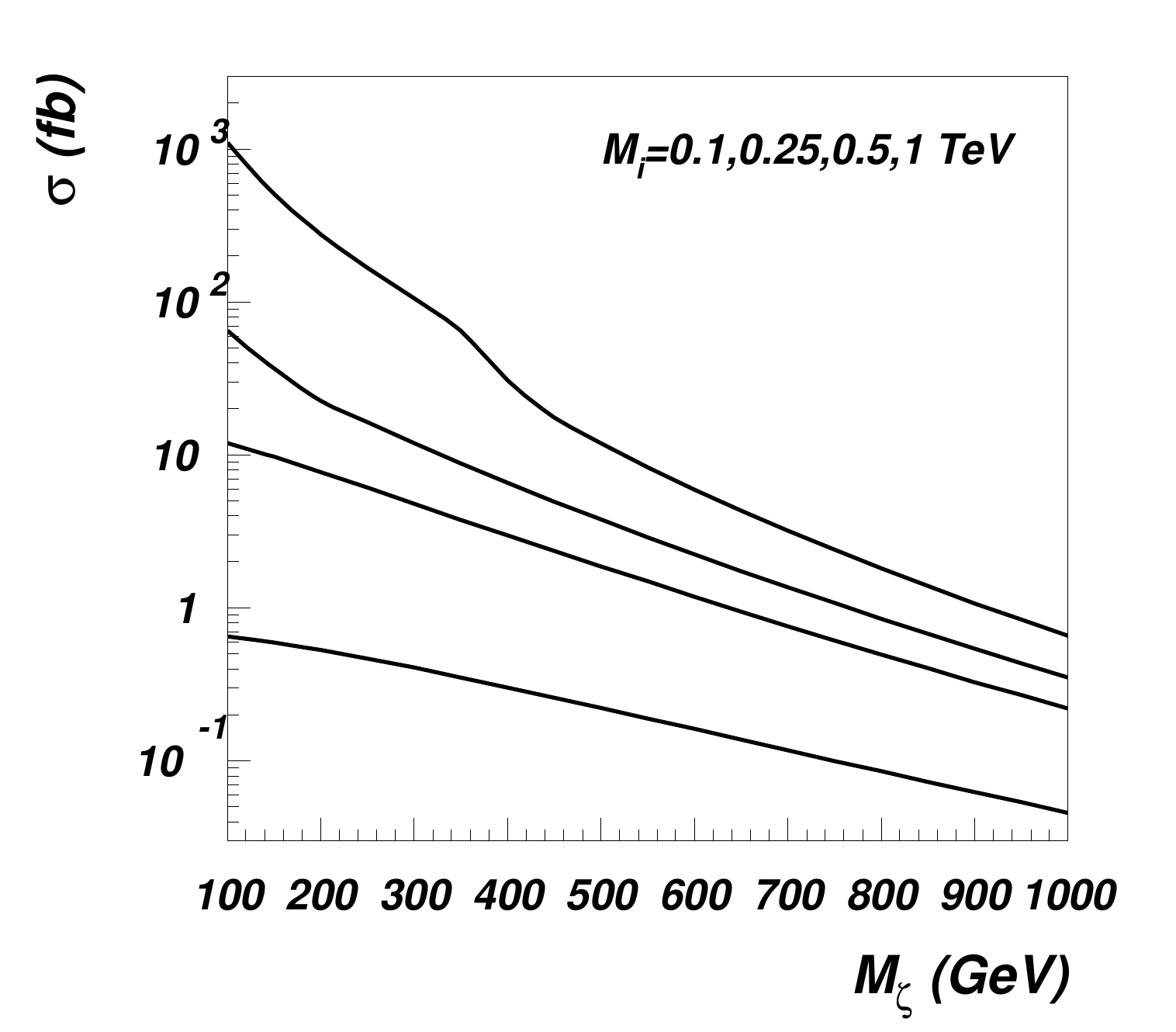}
}
\caption{The LHC production cross section for $pp \to \zeta^+\zeta^-$ (left) and for $pp \to \zeta^+ N_i$ for $M_i=0.1, 0.25, 0.5, 1$ TeV (right), as a function of $M_\zeta$.  The mixing matrix elements are set to unity.  }
\label{fig:heavy e prod}
\end{figure}

\begin{figure}
{\includegraphics[height=6.5cm,width=7.82cm]{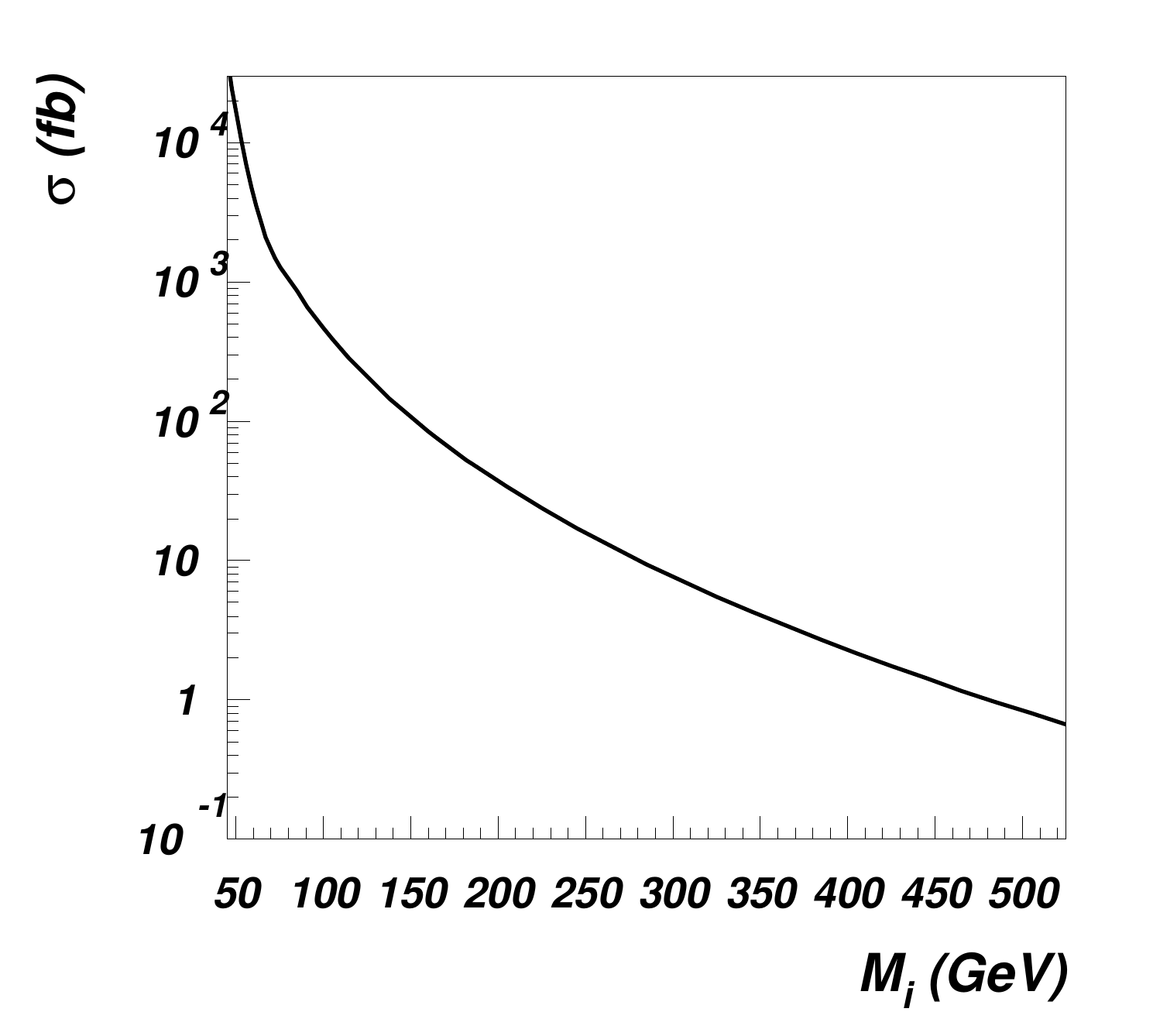}}
\caption{LHC production cross section for $pp\to  N_i N_i $ as a function of $M_i$ assuming unity mixing matrix element.}
\label{fig:heavy n1 prod}
\end{figure}

The heavy leptons may be directly produced through the charged- and neutral current interactions:
\bea
pp &\to& W^\pm/R_{1,2}^\pm \to \zeta^\pm N_i \ ,
\nn
pp &\to& Z/\gamma/R_{1,2}^0 \to \zeta^+ \zeta^- \ , \quad pp \to Z/R_{1,2}^0 \to N_i N_j \ , \quad i,j=1,2 .
\label{Eq:Walking prod}
\eea 
The corresponding Feynman diagrams are given in Fig.~\ref{fig:feynman diagrams}. 
The direct production cross sections for the heavy leptons in MWT are largely independent of the parameters associated to the technicolor sector, see Eq.~(\ref{TCparameters}). Hence, in the direct production of $\zeta^+ \zeta^-$ the only free parameter is the mass of the charged lepton $M_\zeta$. The direct production of $N_i N_j$ and $\zeta^\pm N_i$ depends, in addition to the masses of the leptons, on the $V$ matrix entries of  Eq.~\eqref{Eq: mixing matrix} as follows from Eqs.~\eqref{eq:CCfull} and~\eqref{eq:NCfull}. 
We present in Fig.~\ref{fig:heavy e prod} the LHC cross sections for direct pair and single production of $\zeta$ setting the relevant mixing element to one. The actual production cross section is obtained by multiplying the above result by the square of the corresponding mixing element in the matrix $V$. 
Similarly we present the $N_i N_i$ direct production cross-section in Fig.~\ref{fig:heavy n1 prod}. 

\begin{figure}[h!]
{\includegraphics[height=6.0cm,width=6cm]{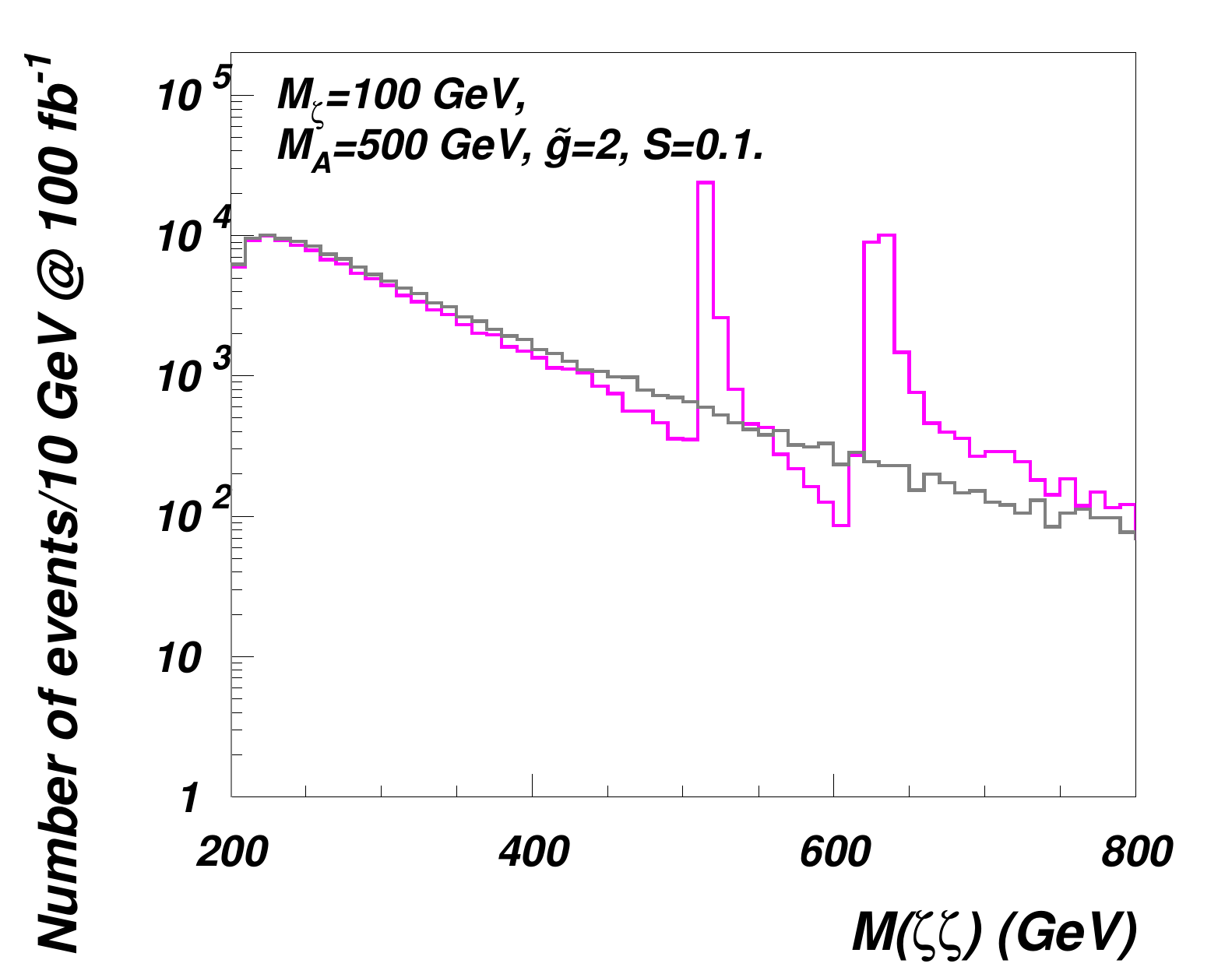}
\hskip 1.1cm
\includegraphics[height=6.0cm,width=6cm]{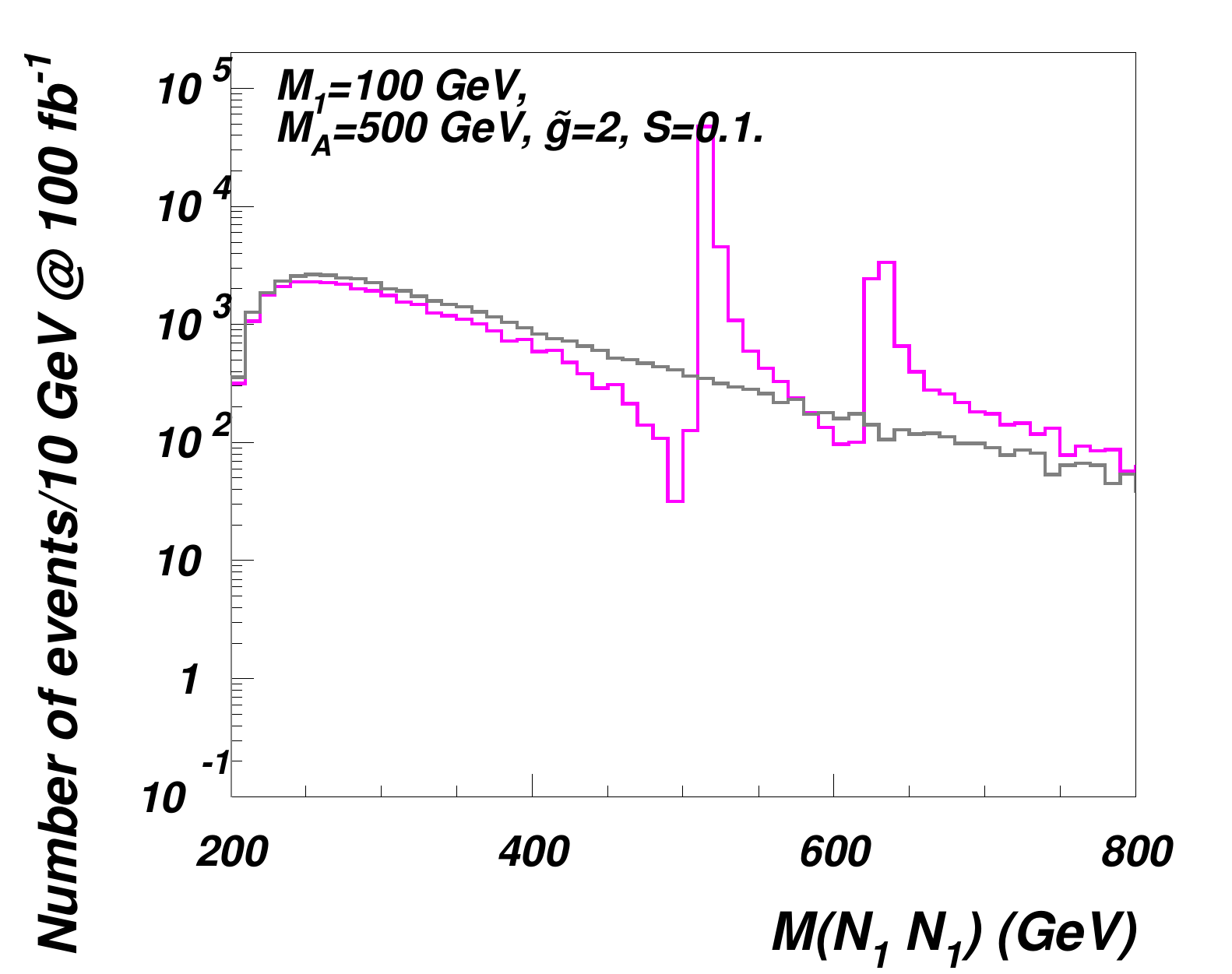}
}
\caption{Invariant mass distributions, $M(\zeta^+ \zeta^-)$ and $M(N_1 N_1)$ in pair production of $\zeta^+ \zeta^-$ (left) and $N_1 N_1$ (right) in the MWT (purple) and when the leptons are added to the SM (grey), assuming unity mixing matrix element. In both frames the first peak corresponds to the $R_1$ resonance while the second corresponds to the $R_2$ resonance. The neutral current coupling of $N_1 N_1$ is axial and therefore more dominated by the $R_1$ resonance. We take the values $\tilde{g}=2, M_A=500, S=0.1$ for the parameters of the technicolor sector. The corresponding masses and widths of $R_{1,2}$ are given in Eq.~(\ref{Eq:Masses and widths of R12})}
\label{fig:Walking direct prod}
\end{figure}

 The final state distributions arising from the direct production of the leptons depend on the specific parameters of the technicolor sector. In particular $R_1$  is a (mostly) axial-resonance 
and $R_2$ is a (mostly)vector-resonance, so $R_1$ mixes mostly with the Z boson while $R_2$ mixes significantly 
with the photon. Consequently the invariant mass distribution of the heavy neutral leptons $N_i N_i$ will be relatively more dominated by the $R_1$ resonance compared to the charged leptons $\zeta^+ \zeta^-$. This is demonstrated in Fig.~\ref{fig:Walking direct prod}. The general form of the mixing in the vector sector is given in Eq.~(\ref{vector mixing}). The masses and widths of $R_{1,2}$ as a function of $M_A, \tilde{g}$ and $S$ are given for two parameter space points in Eq.~(\ref{Eq:Masses and widths of R12}) and studied in general in \cite{Belyaev:2008yj}. 

Production of the heavy leptons can also proceed via the Yukawa-type couplings to the composite Higgs following from Eq.~\eqref{Eq:Yukawa's}. We are following reference \cite{Foadi:2007ue} for an effective way to give masses to the SM fermions in the MWT setup. The composite Higgs may itself be produced through either gluon fusion, vector boson fusion or in association with a SM vector boson:
\bea
gg \to H \ , \quad pp \to q q' H \ , \quad  pp \to H Z / H W     \ .
\eea
The process $gg \to H \to N_i N_i \to  WW\mu\mu $ (within the SM framework) was recently considered in \cite{CuhadarDonszelmann:2008jp} where also a 4th generation of quarks were included that enhance the $gg\to H$ cross section \cite{Kribs:2007nz}, for a recent review of the scenario in which the new leptons are accompanied by a fourth generation of quarks, see \cite{Holdom:2009rf}.
 In the MWT class of models this channel is not expected to be enhanced compared to the SM since the techniquarks are not colored.

On the other hand the associate production of the composite Higgs can be enhanced in these technicolor models \cite{Belyaev:2008yj,Zerwekh:2005wh}. In particular the $pp \to H Z/ H W$ channel can be enhanced by the presence of a light axial-vector resonance as shown in \cite{Belyaev:2008yj}. In that study it was also shown that the vector boson fusion production of the composite Higgs is not expected to be enhanced compared to the SM. We will therefore focus, in the collider signature section, on the associate production of the Higgs.

\begin{figure}
{\includegraphics[height=6.5cm,width=7.82cm]{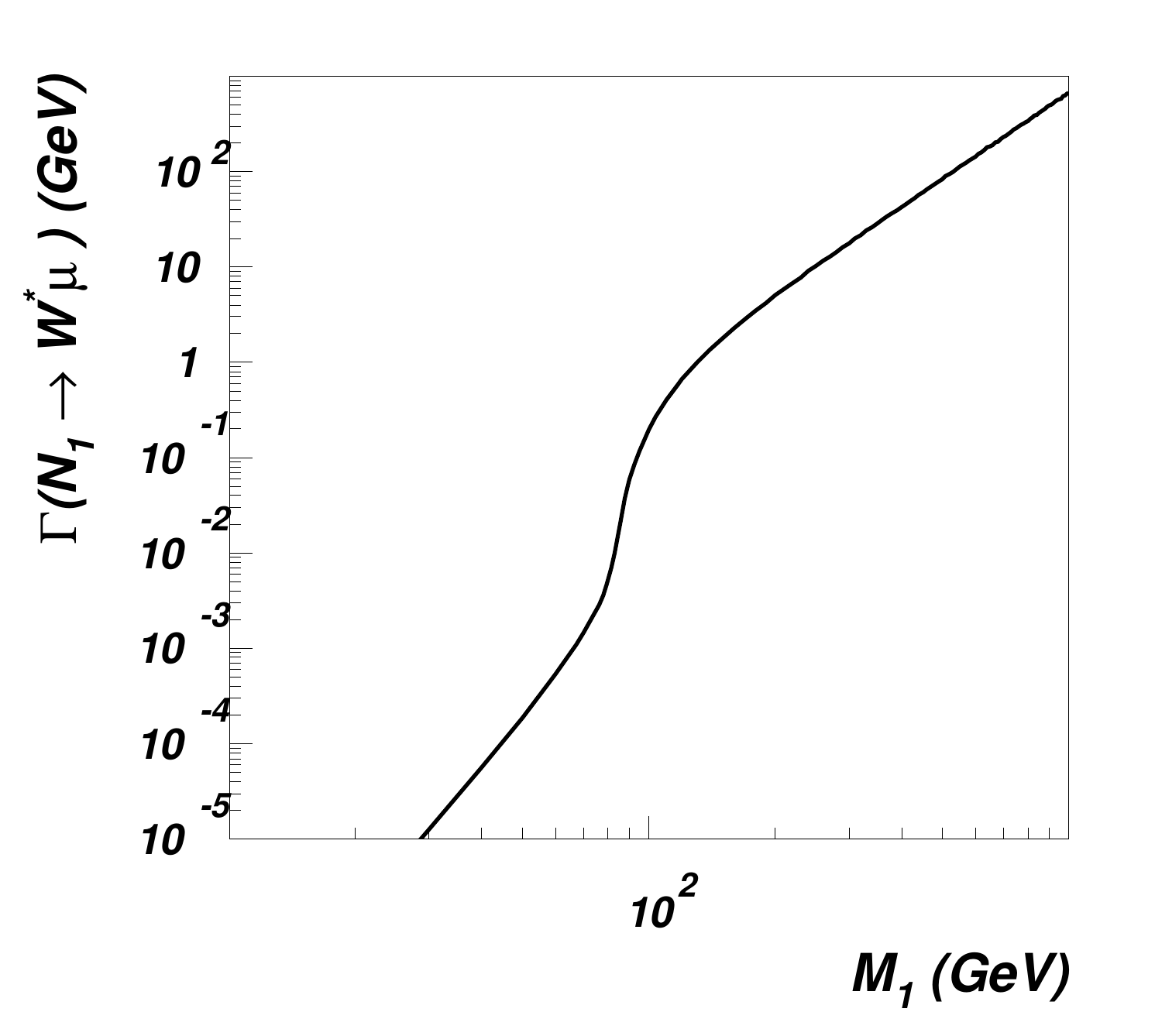}
\includegraphics[height=6.5cm,width=7.82cm]{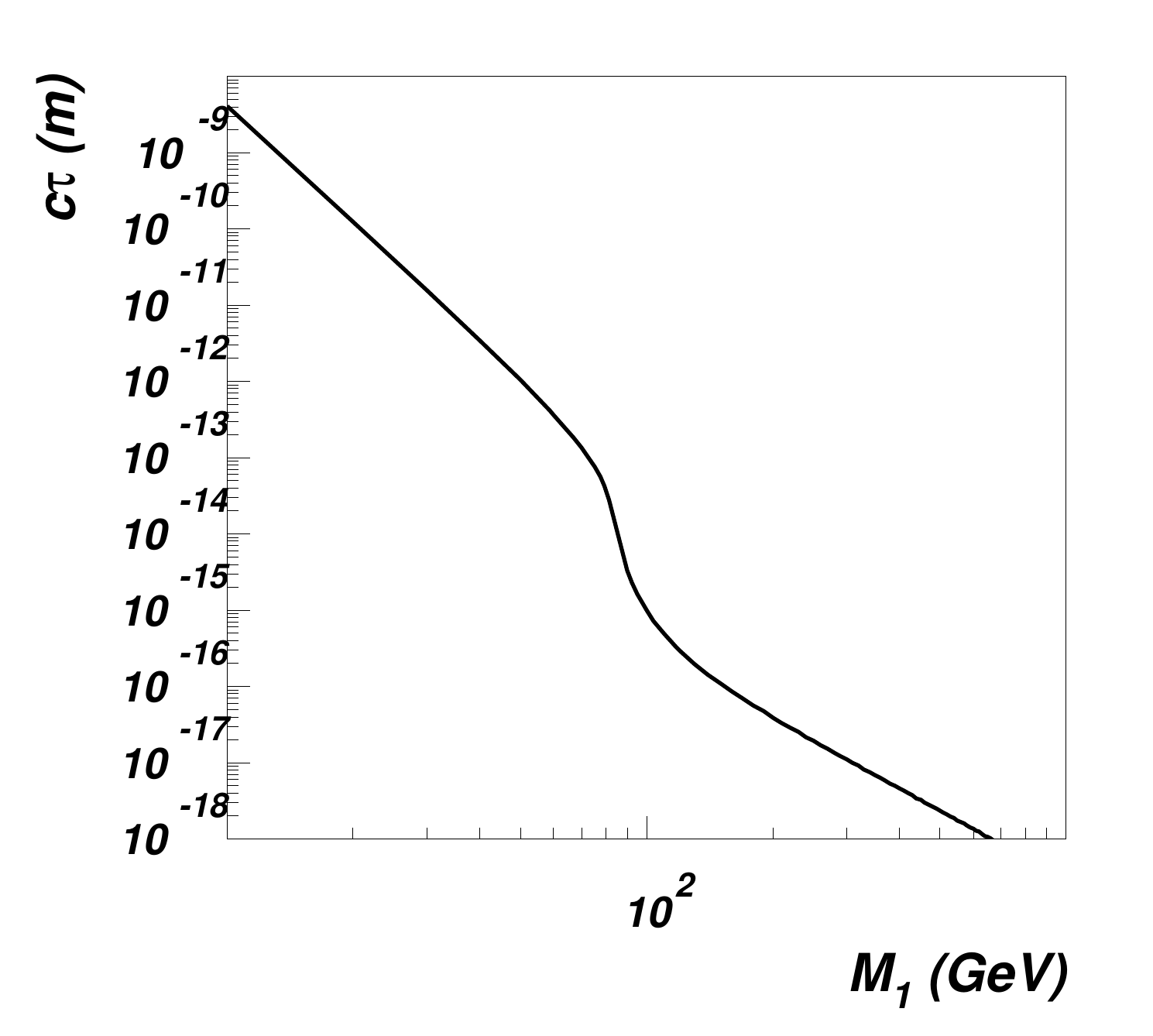}
}
\caption{The width $\Gamma$ (left) and decay length $c \tau$ (right) of $N_1$ assuming decays into $W^* \mu$ with unity matrix element. To obtain the real width the curve in the left frame should be multiplied by the sum of mixing matrix elements squared while the real decay length is found by dividing the curve in the right frame by the same quantity.}
\label{fig:width and time}
\end{figure}

\bigskip
We give expressions for the decay widths of the heavy leptons in Appendix \ref{decays}. The decay patterns depend on the mass hierarchy and the mixings between the leptons. The branching ratio for $N_1$ is particularly simple in the regime where $M_\zeta > M_1$ and $M_H > M_1$. In this case either $N_1$ is absolutely stable, or if it mixes with the SM 
\bea
\Sigma_\ell \rm{Br}(N_1 \to \ell^- W^+)= \Sigma_\ell \rm{Br}(N_1 \to \ell^+ W^-)=0.5.
\label{Eq:branching}
\eea
%This is different from most studies of singlet heavy neutrino additions to the SM, e.g. \cite{del Aguila:2007em,Han:2006ip,Atre:2009rg} since there the heavy neutrino can decay into $Z \nu_\ell$ and to $H \nu_\ell$. 
%The former decay mode is absent in our scenario due to the presence of the lepton doublet in Eq.~\ref{eq:NCfull}. The latter decay mode is also possible in our case but it is more likely to be closed kinematically since the composite Higgs is expected to be heavier than a SM Higgs boson.
Assuming Eq.\eqref{Eq:branching} we plot the decay width and lifetime of $N_1$ in Fig.~\ref{fig:width and time}. This may be compared with the results in \cite{Atre:2009rg}. 
%***
%To understand the size of the mixing parameters that will allow $N_1$ to escape through the detector without 
%decaying or to produce displaced vertices consider its width and lifetime as given in fig.~\ref{fig:width and time}. 
%In this case a value of $\Sigma_\ell |V_{\ell N}|^2\sim 10^{-12}$, which in our case where we assume mixing only with one 
%
We note that a value of $\sin\theta^{\prime} \sim 10^{-6}$ would yield a decay length of $\sim$ 1 m which is enough for a relativistically boosted particle to escape detection at the LHC and be considered as missing energy in the various processes \cite{Basso:2008iv}. 
%If we additionally take into account that the $N_1$ neutrino produced in the MWT model, is typically boosted by a factor 
%$\sim \frac{M_{R_{1,2}}}{M_1}$, where $R_{1,2}$ are the new heavy vector meson mass eigenstates discussed below Eq.~(\ref{Eq:covderivM}) this could allow the neutrino to escape as pointed out in a different model in \cite{Basso:2008iv}.  
%An example of hierarchy realizing a lifetime of this order would be $\tan \theta' \sim \frac{m_e}{m_t}$.
%***

\subsection{Collider signatures of heavy leptons with an exact flavor symmetry}
 
Let us first consider the limit in which the new leptons do not mix appreciably with the SM ones.

If $M_\zeta > M_1$ then $N_1$ constitutes a long lived neutral particle and will give rise to missing momentum $\mpt$ and missing energy $\met$ signals. In particular the decay mode $H\to N_1 N_1$ gives rise to an invisible partial width of the composite Higgs.

\begin{figure}[h!]
{\includegraphics[height=2.5cm,width=5cm]{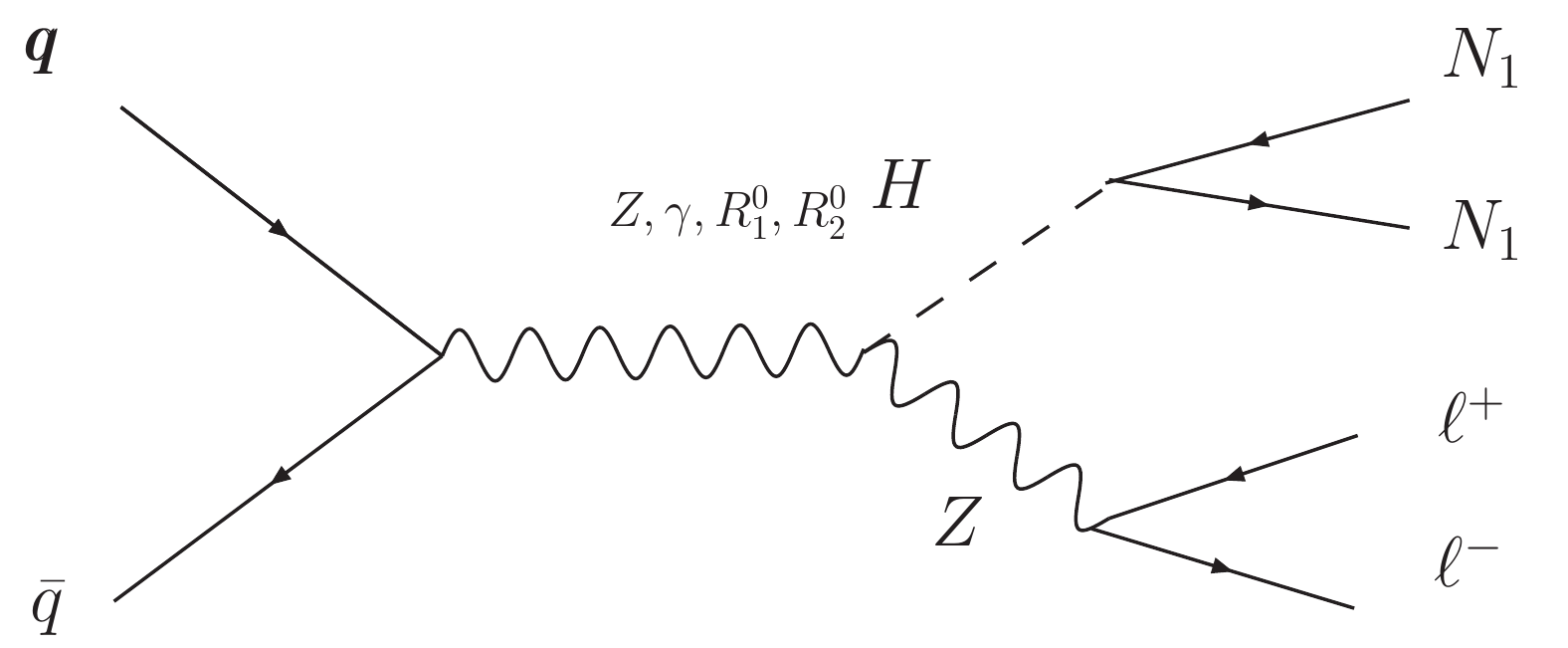}
\includegraphics[height=2.5cm,width=5cm]{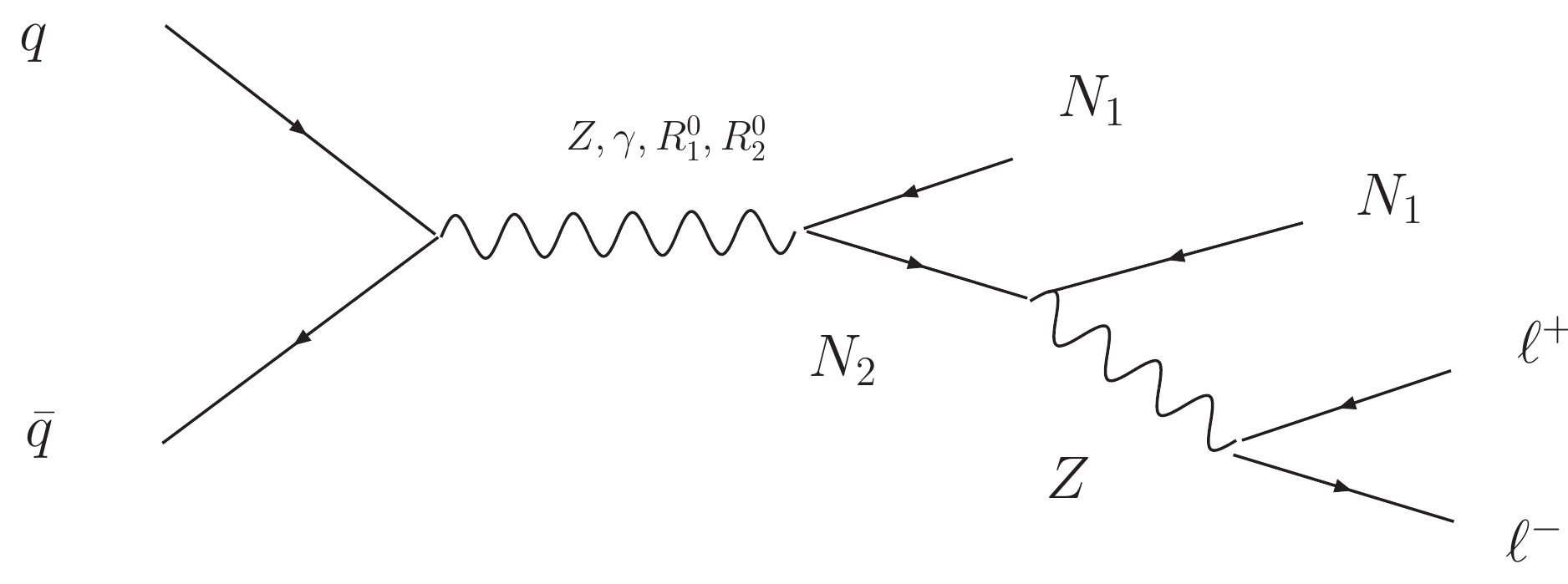}}
\caption{Feynman diagrams for the $\ell^+\ell^- +\mpt$ signal due to heavy leptons in the MWT model}
\label{fig:missEfromH}
\end{figure}

As pointed out in \cite{Belyaev:2008yj,Zerwekh:2005wh}, the cross-section for $ZH$ production can be enhanced in MWT models because the axial-vector resonance can be light  \cite{Foadi:2007ue,Appelquist:1998xf}.
Here the $\ell^+\ell^- +\mpt$ state will receive contributions both from $ZH$ and $N_1, N_2$ production, {\bf as shown in Fig.~\ref{fig:missEfromH}}.
We therefore study the proces $pp \to Z N_1 N_1 \to \ell^+ \ell^- + \mpt$. We consider limiting values of the parameters such that either the Higgs or the $N_2$ state is too heavy to contribute significantly as well as parameters where both contribute in the process. 

 The acceptance cuts relevant for LHC are
\begin{eqnarray}
|\eta^{\ell}|<2.5\ , \quad  p_T^\ell> 10 \mbox{ GeV} \ , \quad   \Delta R(\ell\ell) > 0.4  \ .
\label{eq:cuts1}
\end{eqnarray}
Here $\ell$ is a charged lepton, $\eta^{\ell}$ and $p_T^\ell$ are the pseudo-rapidity and transverse momentum of a single lepton while $\Delta R$ measures the separation between two leptons in the detector. $\Delta R$ is defined via the difference in azimuthal angle $\Delta\phi$ and rapidity $\Delta\eta$ between two leptons as $\Delta R\equiv \sqrt{(\Delta\eta)^2+(\Delta\phi)^2}$. 

The main sources of background come from di-boson production followed by leptonic decays \cite{Godbole:2003it,Davoudiasl:2004aj,Meisel:2006}
\beq
ZZ \to \ell^+\ell^- \nu \bar{\nu} \ , \ W^+ W^- \to \ell^+ \nu \ell^-\bar{\nu} \ , \ ZW \to \ell^+\ell^- \ell \nu 
\eeq
where in the last process the lepton from the W decay is missed.
 
 We impose the additional cuts
\begin{eqnarray}
 |M_{\ell \ell}-M_Z|<10 \mbox{ GeV} \ , \quad {\rm and} \quad \Delta \phi(\ell\ell) < 2.5  \ .
\label{eq:cuts2}
\end{eqnarray}
The first is meant to reduce the WW background by requiring the invariant mass of the lepton pair to reproduce the Z boson mass. The second cut on the azimuthal angle separation together with taking large $\mpt$  reduces potential backgrounds such as single Z production + jets with fake $\mpt$ \cite{Godbole:2003it,Davoudiasl:2004aj}.

\begin{figure}[h!]
{\includegraphics[height=5.5cm,width=6cm]{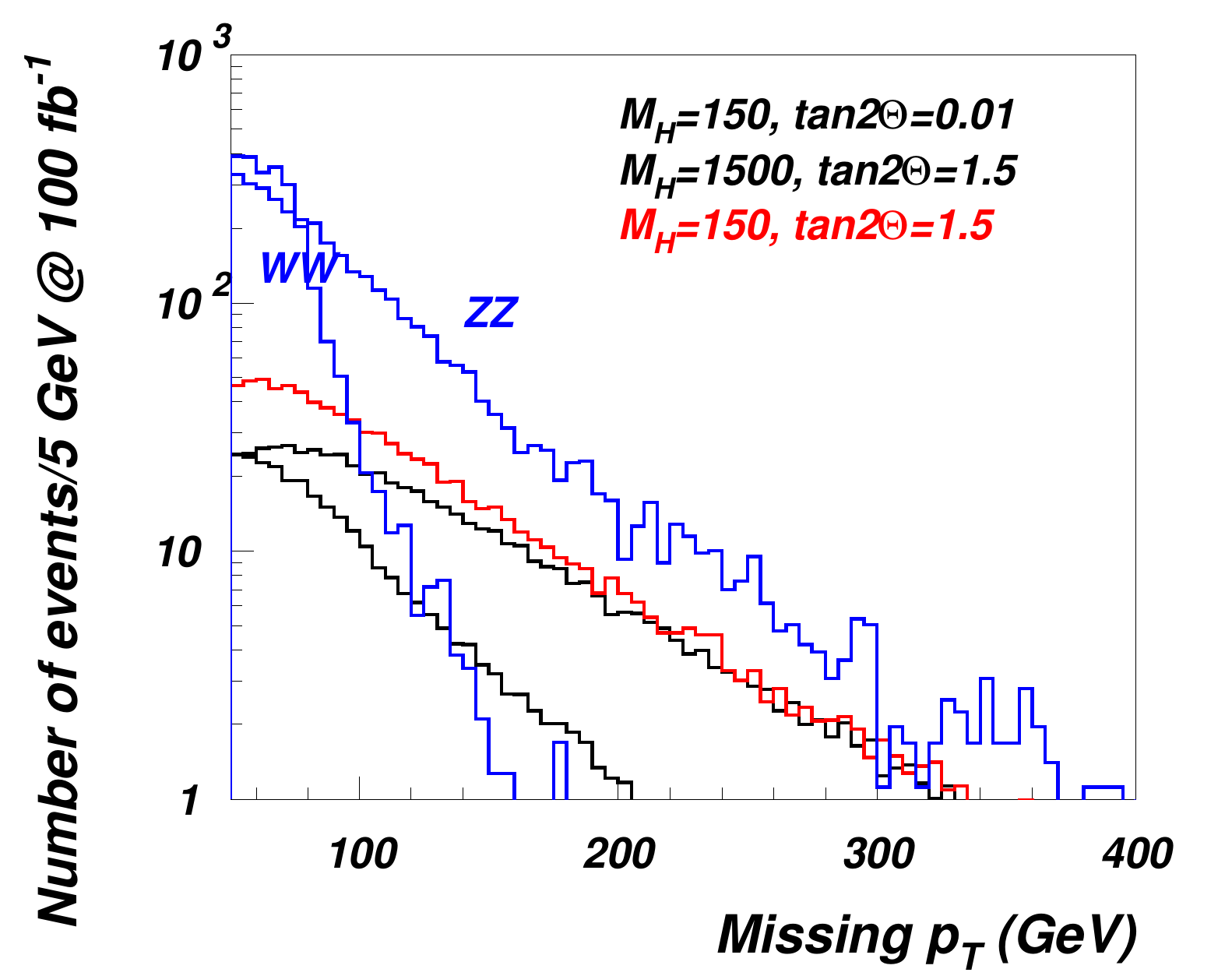}
\hskip 1.1cm
\includegraphics[height=5.5cm,width=6cm]{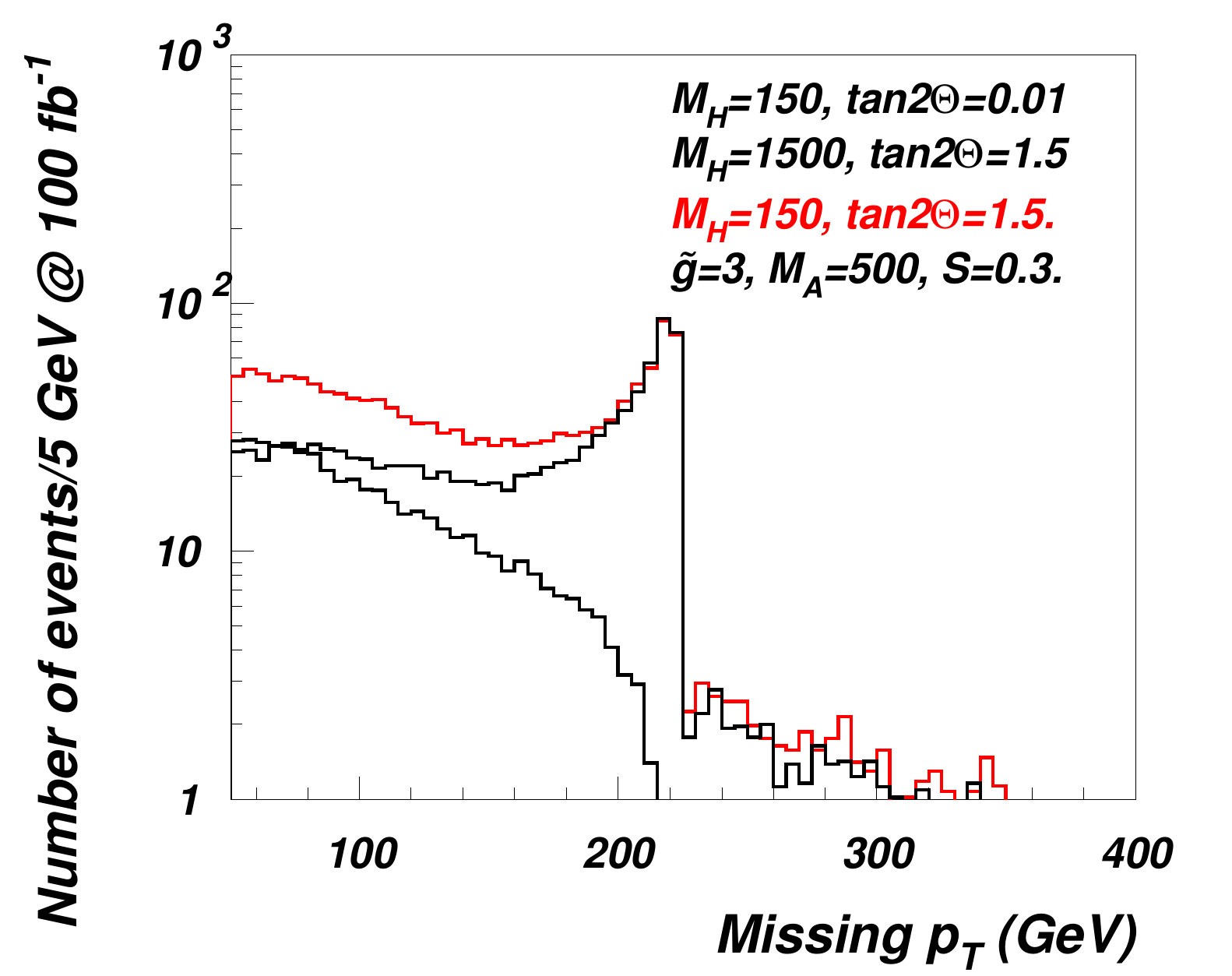}
}
\caption{$\ell^+\ell^- +\mpt$ signal from $pp \to Z N_1 N_1 \to \ell^+ \ell^- N_1 N_1$ { in the SM (left) and in the MWT (right)}. In blue the SM background from ZZ and WW production. Upper black line corresponds to $M_H=150$ GeV, $M_1=50$ GeV, $\tan 2\theta$=0.01 ($M_2=20$ TeV);
Lower black line corresponds to $M_H=1500$ GeV, $M_1=50$ GeV, $\tan 2\theta$=1.5($M_2=175$ GeV);
Red line corresponds to $M_H=150$ GeV, $M_1=50$ GeV, $\tan 2\theta$=1.5($M_2=175$ GeV)}
\label{fig:Walking invisible higgs}
\end{figure}

\begin{figure}[h!]
{\includegraphics[height=6.5cm,width=7.82cm]{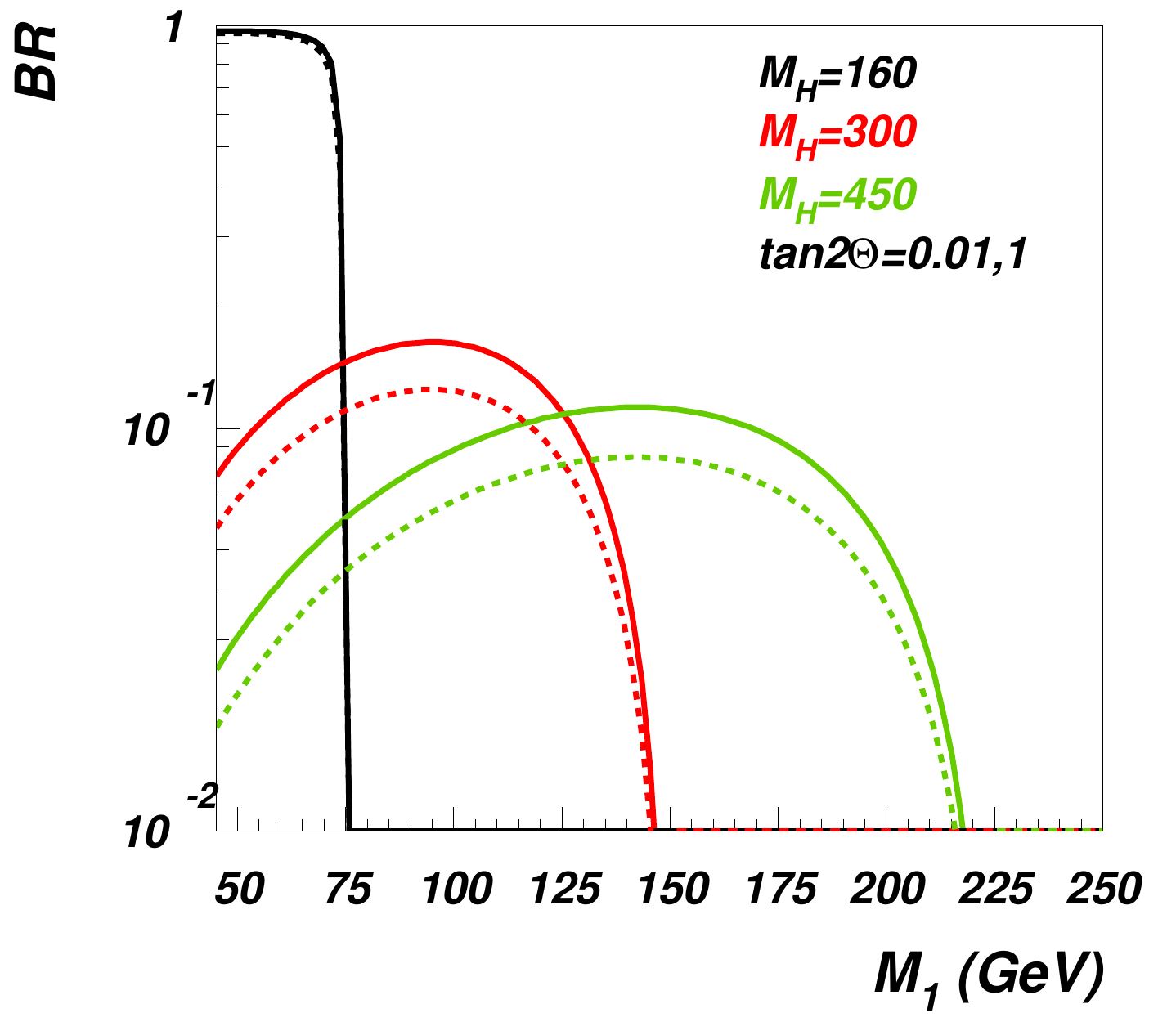}}
\caption{The invisible branching fraction of the composite Higgs $Br[H\to N_1 N_1]$}
\label{fig:branching}
\end{figure}

The results are given in Fig.~\ref{fig:Walking invisible higgs} assuming a fully invisibly decaying Higgs. { On the left panel we show the signal and background arising from the SM featuring the new heavy leptons.} On the right hand panel we show the same signal but in the MWT model. For $\mpt > 100$ GeV where the signal could potentially be observed, the Higgs production channel dominates and in the MWT model a very distinct $\mpt$ distribution arises due to the effect of the $R_1$ resonance. While invisible decays of a SM model Higgs at most appear as an excess of events compared to the background in e.g. $\mpt$ distributions, the presence of a light axial-vector resonance results in a peaked distribution, different from the shape of the background, making it a more striking signal. This was also found in the context of Higgs decays to technibaryon Dark Matter candidates in \cite{Foadi:2008qv}. The peak will degrade for smaller values of $\tilde{g}$.
However, for a light axial resonance a relatively large value of $\tilde{g}$ is favored by unitarity arguments \cite{Foadi:2008xj,Foadi:2008ci} and electroweak precision observables \cite{Belyaev:2008yj,Foadi:2007ue,Foadi:2007se}
The peak will also degrade and move to higher $\mpt$ for larger mass of the axial resonance.    
Also, in Fig.~\ref{fig:Walking invisible higgs} we have assumed a fully invisibly decaying Higgs. The true invisible branching fraction of the Higgs will depend on the mass of the $N_1$ particle as shown in Fig.~\ref{fig:branching}. 

An invisible partial width of the composite Higgs has been searched for at LEP in the proces $e^+ e^- \to HZ$ with $Z$ decaying hadronically \cite{Abbiendi:2006gd}. However, no limits were achieved for $M_H>114 GeV$. 
The same final state from $HZ/HW$ production in $p p$ collisions has been considered by several authors, also  recently in the context of technicolor models \cite{Foadi:2008qv}. See \cite{Foadi:2008qv} for references to the literature.
The LHC discovery potential for an invisibly decaying Higgs in this final state at LHC has also been investigated at detector level \cite{Meisel:2006,Gagnon:2005}. It was found that the $HW$ mode is not promising \cite{Gagnon:2005} while  the ZH mode remains challenging. 
With a SM production cross section of $HZ$ a significance of 3.43 $\sigma$ was achieved at $M_H=160$ Gev dropping to a 2 $\sigma$ excess at $M_H=200$ GeV \cite{Meisel:2006}.

{Thanks to the possible increase in the $HZ$ production cross section found in \cite{Belyaev:2008yj} together with the resonance like structure in the $\mpt$ distribution displayed in Fig.~\ref{fig:Walking invisible higgs}, we believe that this channel could be interesting to investigate the interplay between new long-lived heavy neutrinos and composite vector states at LHC. }

 \bigskip
\bigskip
If instead $M_\zeta < M_1$ then $\zeta$ can be a long-lived CHAMP (Charged Massive Particle). Collider signatures of long lived charged leptons could either be displaced vertices or, if the charged lepton 
decays outside the detector, a muon like signal for which the heavy mass should be reconstructable. 
%This will also be the case if $M_1> M_\zeta$ provided the mass-splitting is sufficiently small. 
Such a long-lived CHAMP arises in several scenarios and has been study in some detail, for a review see e.g. \cite{Fairbairn:2006gg}.

%To quantify the interplay between the heavy vectors and heavy leptons we compute the ratio of couplings $\frac{g_{N_1^2 R_1}^2}{g_{N_1^2 R_2}^2}$ and $\frac{g_{R_1 \zeta \zeta }^2}{g_{R_2 \zeta \zeta }^2}$ where
%\bea
%g_{N_1^2 R_i}^2 \equiv \lim_{M_1\to 0} \frac{1}{M_{R_i}^2}|\mathcal{A}(R_i \to N_1 N_1)|^2=
%\frac{2}{3}\left(g_2 N_{2(1+i)}-g_1 N_{1(1+i)}\right){}^2 \ ,
%\eea  
%and similar for the other states.
%We plot the ratios in Fig~\ref{fig:couplings}.
%\begin{figure}[h!]
%{\includegraphics[height=7.0cm,width=7cm]{figures/RNNcouplings}
%\hskip 1.1cm
%\includegraphics[height=7.0cm,width=7cm]{figures/REEcouplings}
%}
%\caption{Ratios of the couplings of $R_i$ to the heavy leptons. $R_1$ couples predominantly to the neutral leptons while $R_2$ couples predominantly to the charged leptons.}
%\label{fig:couplings}
%\end{figure}

In \cite{Allanach:2001sd} a Herwig based study of the LHC reach for long-lived leptonic CHAMPs was considered, based on a direct production cross-section identical to that in Fig.~\ref{fig:heavy e prod}. With a discovery criterion of 5 pairs of reconstructed opposite charge heavy leptons a reach of $M_\zeta =950$ GeV at $100 \textrm{fb}^{-1}$ was found, reduced to 800 GeV without specialized triggers. We can expect this reach to be improved in our model by searching also for single $\zeta$ production channel in Fig.~\ref{fig:heavy e prod}. Additionally it was found that long-lived leptonic and scalar CHAMPS could be dstinguished for masses up to 580 GeV.

The discovery potential for long-lived CHAMPS has also been studied at detector level for LHC. The CMS and Atlas collaboration has considered various long-lived CHAMPS \cite{Giagu:2008im}. From their results we infer that 3 signal events with less than one background event could be observed in CMS with 1 $\textrm{fb}^{-1}$ and $M_\zeta \sim 300$ GeV and similar in Atlas. More precisely in \cite{Giagu:2008im} 3 signal events could be seen in direct pair production of 300 GeV KK taus with a pair production cross-section of 20 fb, very similar to what we find in Fig.~\ref{fig:heavy e prod}.    
{ Figure \ref{fig:Walking direct prod} shows that it is interesting to investigate the invariant mass distribution of the leptonic CHAMP. }

\subsection{Collider Signatures of Promiscuous Heavy Leptons}
\begin{figure}[h!]
{
\includegraphics[height=3cm,width=5cm]{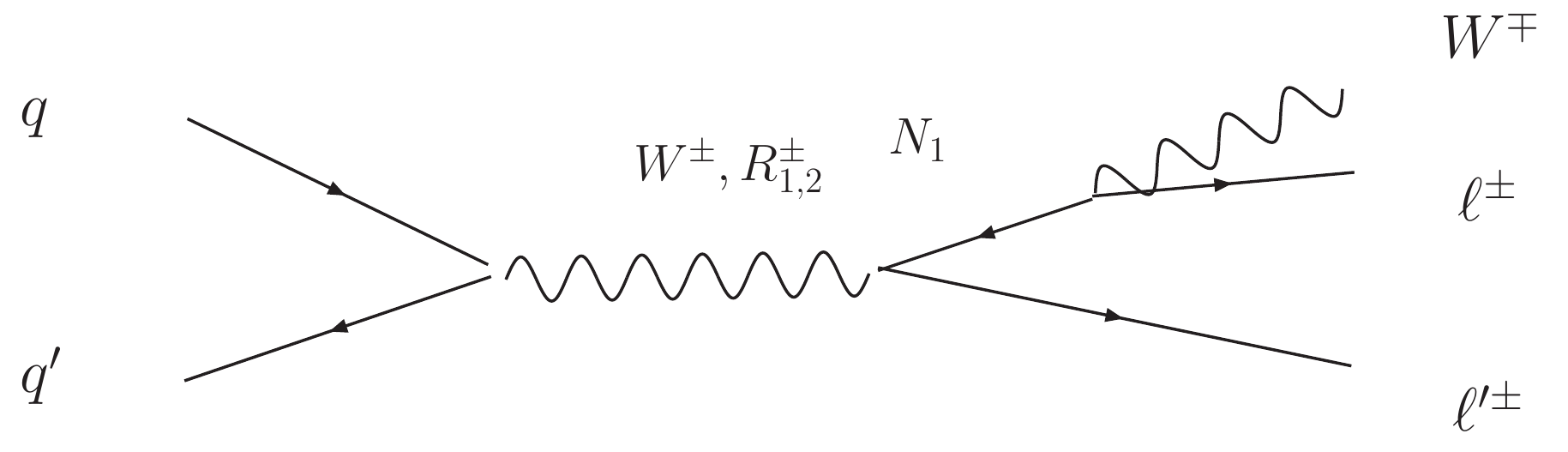}
\includegraphics[height=3cm,width=5cm]{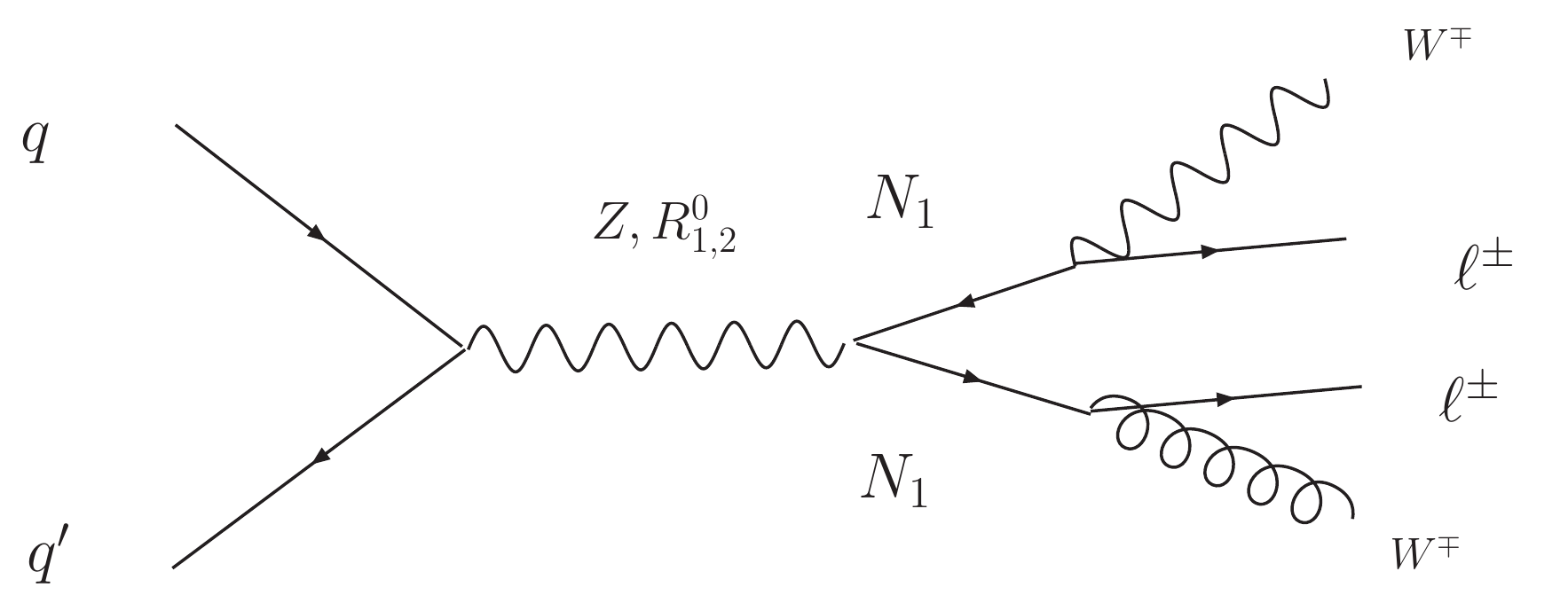}
}
\caption{Same sign leptons from production of $N_i$. We will consider the case where the W's decay to jets. Such that the final states we consider are $\mu^\pm \mu^\pm jj$ (left) and $\mu^\pm \mu^\pm jjjj$ (right)}
\label{fig:samesign}
\end{figure}

If the heavy leptons mix with the SM leptons, this will give rise to Lepton Number Violating (LNV) processes with same sign leptons and jets in the final state, e.g.
\bea
pp &\to& W^\pm/R_{1,2}^\pm \to \mu^\pm N_i \to  \mu^\pm  \mu^\pm W^\mp \to \mu^\pm \mu^\pm jj
\nn
pp &\to& Z/R_{1,2}^0 \to N_i N_j \to \mu^\pm  \mu^\pm W^\mp  W^\mp \to \mu^\pm \mu^\pm jjjj \ , \quad i=1,2 ,
\label{Eq:mumujjjj}
\eea 
as in Fig.~\ref{fig:samesign}.

\begin{figure}[h!]
{
\includegraphics[height=6.5cm,width=8cm]{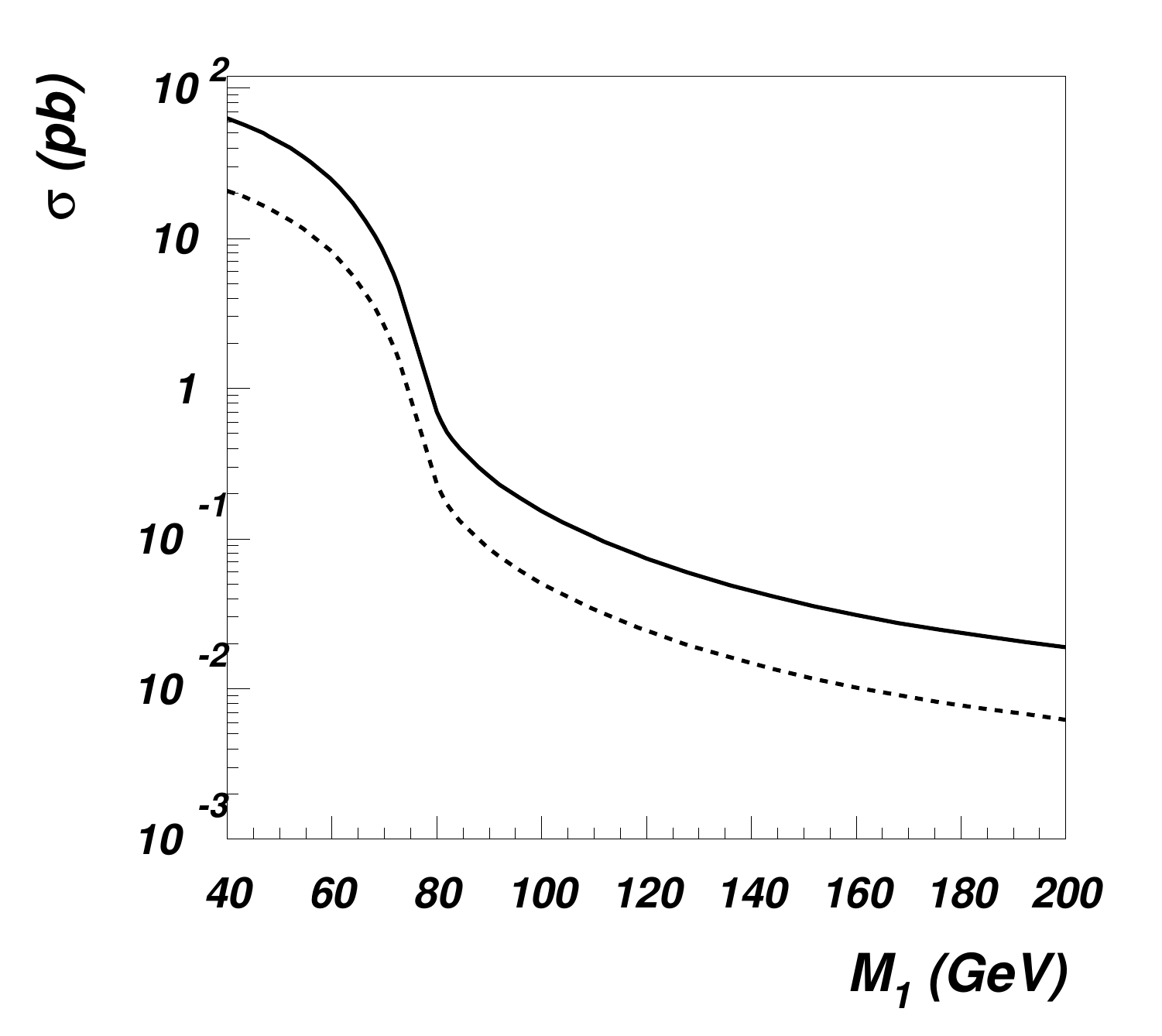}
}
\caption{LHC cross-sections: $\sigma(p p \to N_1 \mu^\pm)$ (solid black) and $\sigma(p p \to \mu^\pm \mu^\pm jj)$ (dotted black) without imposing any cuts. We take $\sin\theta'=0.098$ and $\cos\theta =1$ }
\label{like sign leptons}
\end{figure}

The production cross section $pp \to N_1 N_1 \to \mu^\pm \mu^\pm jjjj$ may be inferred from Fig.~\ref{fig:heavy n1 prod} using the branching in Eq.~(\ref{Eq:branching}) (assuming $M_\zeta > M_1$ and $M_H > M_1$). The production cross-section for $pp\to \mu^\pm \mu^\pm jj$ is given in in Fig.~\ref{like sign leptons}.
To compare with  \cite{del Aguila:2007em} we have taken  $\cos\theta \sin\theta'=0.098$ and $\tan 2\theta =0$ such that $N_2$ is decoupled. More generally the contribution from $N_2$ will be supressed both by the heavier mass and the smaller mixing matrix element. 
%(Note again two differences between our scenario and that of \cite{del Aguila:2007em,Han:2006ip,Atre:2009rg}. Due to the presence of the left-handed neutrino doublet we have $Br(N_i \to Z \nu)=0$ from Eq.~\ref{eq:NCfull}. With only right-handed sterile neutrinos this branching would be $\frac{1}{4}$. Similarly we have a smaller branching $N_i \to H\nu$ because the mass of the composite Higgs is larger than that of a SM-like Higgs.) 

The potential for observing the $\mu^\pm \mu^\pm jj$ final state has been extensively studied in scenarios with heavy right-handed neutrino singlets, both in the $3\nu$-SM \cite{del Aguila:2005pf,del Aguila:2006dx,Han:2006ip,Atre:2009rg} and in the presence of additional new physics \cite{delAguila:2007ua,Basso:2008iv}. Same sign lepton final states have been searched for at the Tevatron in \cite{Abulencia:2007rd, Aaltonen:2008vt}. The $pp \to W^\pm \to \ell^\pm N_i \to \ell^\pm \ell'^\pm j j$ process in the $3\nu$-SM was studied in \cite{del Aguila:2007em} at the level of a fast detector simulation. While backgrounds for same sign lepton production are smaller than for opposite sign lepton production, arising in the Dirac limit, they were found to be significantly larger than previously estimated in parton level processes, in particular for $M_i<M_W$.   

Again the production cross-sections are largely unaffected by the presence of heavy vectors and is as shown in Fig. \ref{like sign leptons}. However the shape of the distributions are affected by the presence of the heavy vectors.

To study these processes we impose jet acceptance cuts in addition to the leptonic acceptance cuts given in Eq.~(\ref{eq:cuts1})
\begin{eqnarray}
|\eta^{j}|<3\ , \quad  p_T^j> 20 \mbox{ GeV} \ , \quad   \Delta R(\ell j) > 0.5  \ .
\label{eq:cuts3}
\end{eqnarray}

The resulting invariant mass distributions for $\mu^-\mu^- jj$ (left) and $\mu^-\mu^- jjjj$ (right) are given below in Fig.~\ref{fig:Walking direct production}. We have again taken $\cos\theta=1 , \sin\theta'=0.098$. While $\sin\theta'$ determines the mixing between $N_1$ and $\ell$ and therefore is constrained by experiment, $\cos\theta$ is not. This means that the production cross section of $N_1, N_1$ potentially is significantly larger than the $N_1 \mu$ production cross-section, compare Figs.~\ref{fig:heavy n1 prod} (left) and~\ref{like sign leptons}. If at the same time $N_1$ only decays to $W \mu$ as in Eq.~(\ref{Eq:branching}) with $Br(N_1 \to W^+ \mu^-)=\frac{1}{2}$ as the only mode we find the result given in the right frame of Fig.~\ref{fig:Walking direct production}  
\begin{figure}[h!]
{\includegraphics[height=6.5cm,width=7cm]{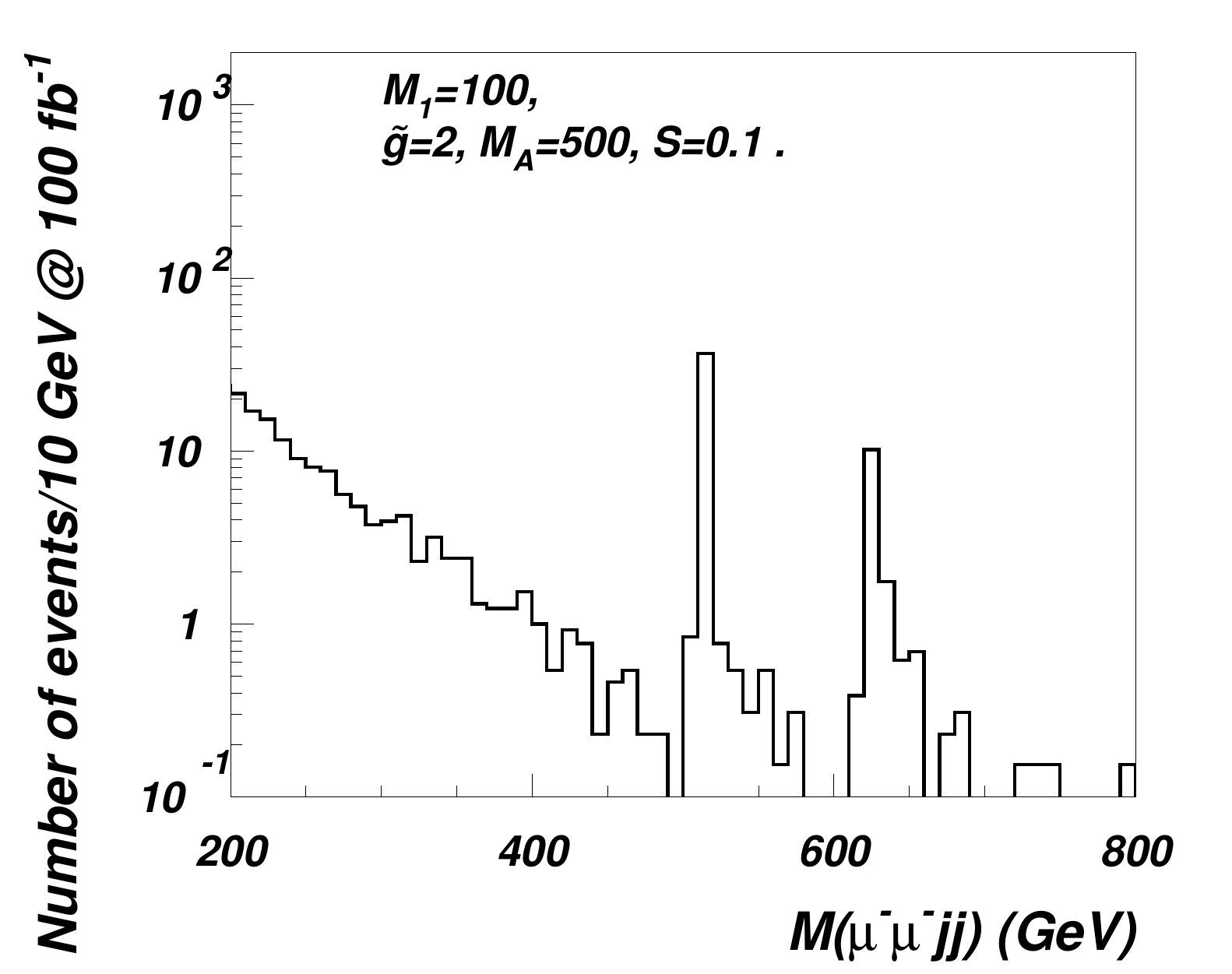}
\hskip 1.1cm
\includegraphics[height=6.5cm,width=7cm]{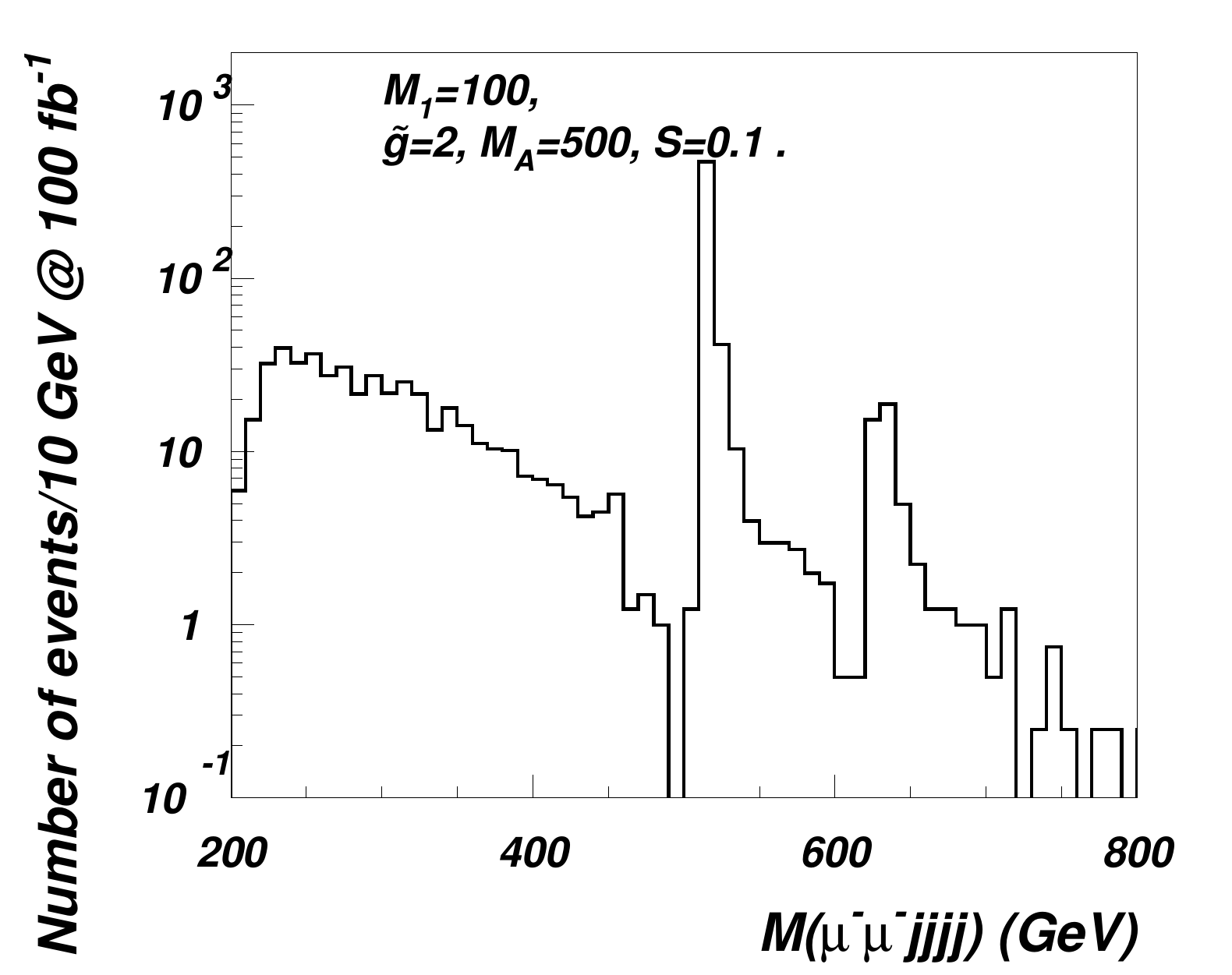}
}
\caption{Invariant mass distributions: $pp \to N_1 \mu^- \to \mu^-\mu^- jj $ (left) and $pp \to N_1 N_1 \to \mu^-\mu^- jjjj$ (right). The parameters in the technicolor sector are fixed to be $M_A=500$ GeV, $S=0.1, \tilde{g}=2$ while the new lepton sector parameters are $M_1=100 GeV, \cos\theta=1, \sin\theta'=0.098$. $N_2$ is decoupled for these parameters.}
\label{fig:Walking direct production}
\end{figure}

The above shows that the interplay of heavy neutrinos and composite vector mesons can lead to striking signatures at the LHC.

%%%%%%%%%%%%%%%%%%%%%%%%%%%%%%%%%%%%%%%%%%%%%%%%%%%%%%%%%%%%%%%
\section{An intriguing structure }
We argued that linking the 4th lepton family to a technicolor sector rather than to a fourth family of ordinary quarks  renders its presence at the electroweak scale natural. What is even more interesting is that this link makes the new leptons much less conventional than a sequential type fourth generation. If a grand unification mechanism is at play it is natural to expect that at some energy scale (higher than the weak scale) the weak group extends to \cite{Gudnason:2006mk}: 
\begin{equation}
SU_1(2) \times U_1(1) \times SU_2(2) \times U_2(1) \ .
\end{equation}  
The SM leptons and quarks are uncharged under the first copy of $SU(2)\times U(1)$ while they have the standard charges under the second copy. The fourth family of leptons has the SM like assignment under the first copy but is uncharged under the second copy. We reproduce the table of charges from \cite{Gudnason:2006mk} below:
\begin{table}[h]
\caption{MWT + One SM Family enlarged gauge group}
\begin{center}
\begin{tabular}{c|c|c|c|c|c|c}
&$SO_{TC}(3)$&$SU_1(2)$&$U_1(1)$&$SU_c(3)$&$SU_2(2)$&$U_{2}(1)$ \\
\hline \hline
$q_L$ &1&1&0&3&2&1/6 \\
$u_R$& 1&1&0 &3 &1 & 2/3 \\
$d_R$ &1&1& 0&3 &1&-1/3\\
$L $&1&1 &0&1&2&-1/2 \\
$e_R$ &1&1& 0 &1&1&-1 \\
\hline
$Q_L$ &3&2&1/6&1&1&0 \\
$U_R$ & 3 &1 & 2/3&1&1&0 \\
$D_R$ & 3&1 &-1/3&1&1&0\\
${\cal L}_L $& 1&2&-1/2&1&1&0 \\
$\zeta_R$ & 1 &1&-1&1&1&0 \\
\end{tabular}
\end{center}
\label{tabledouble}
\end{table}
This assignment allows to arrange the low energy matter  fields
into complete representations of $SU(5)\times SU(5)$. 
We summarize in table \ref{GUT} the technicolor and SM fermions
transformation properties with respect to the grand unified group.
\begin{table}[h]
\caption{GUT}
\begin{center}
\begin{tabular}{c|c|c}
&$SU(5)$&$SU(5)$ \\
\hline \hline
$\bar{A}_{SM}$ &1&$\overline{10}$\\
$F_{SM}$ & 1&5 \\
\hline
$\bar{A}_{MWT}$ &$\overline{10}$&1\\
$F_{MWT}$& 5&1 \\
\end{tabular}
\end{center}
\label{GUT}
\end{table}
Here the fields $A$ and $F$ are standard Weyl fermions and the gauge
couplings of the two $SU(5)$ groups need to be the same. This is a minimal embedding and
others can be envisioned.

We further require the extended gauge group to spontaneously break to a single $SU(2)\times U(1)$ which we then identify with the SM weak and hypercharge gauge group. We do not speculate on the mechanism behind the breaking of the extended group and immediately introduce a two by two complex field $\widetilde{H}$ transforming with respect to the extended group as follows:
\begin{equation}
\widetilde{H} \qquad (2,1/2,2,-1/2) \ . 
\end{equation} 
The covariant derivative is:
\begin{equation}
D\widetilde{H} = \partial\widetilde{H} - i\,g_1\,W_1\widetilde{H} - i \frac{1}{2}\, g^{\prime}_1\,B_1\widetilde{H}+ i\,g_2 \, \widetilde{H} W_2 + i \frac{1}{2}\, g^{\prime}_2 \, B_2 \widetilde{H}
\end{equation}
where we have suppressed the Lorentz indices and 
\begin{equation}
W_{1/2} = W_{1/2}^a \frac{\sigma^a}{2 }
\end{equation}
are the intermediate massless gauge bosons for the two non abelian SU(2) groups while the $B$s are the gauge bosons for the abelian part. $\sigma^a$ are the Pauli matrices normalized according to ${\rm Tr}[\sigma^a \sigma^b] = 2\delta^{ab}$.

When this scalar field acquires a diagonal nonzero vacuum expectation value
\begin{equation}
<\widetilde{H}>  = \tilde{v} \, {\mathbf 1} \end{equation} then the kinetic term for $\widetilde{H}$ evaluated on the vacuum reads:
\begin{equation}
{\rm Tr}\left[ D\widetilde{H}^{\dagger} \ D\widetilde{H}\right] \rightarrow 
\frac{\tilde{v}^2}{2}\left[ (g^{\prime}_1B_1 -g^{\prime}_2B_2)^2 +\sum_{a=1}^3 (g_1W_1^a -g_2W_2^a)^2 \right] \ .
 \end{equation}
 We identify four massive states :
 \begin{equation}
 B^{\prime} = \cos \beta\, B_1 - \sin \beta \, B_2 \ , \qquad {W^{\prime}}^a =    \cos \omega \, W_1^a - \sin \omega \, W_2^a 
 \end{equation}
 with mass
 \begin{eqnarray} 
 M_{B^{\prime}} = \tilde{v} \sqrt{{g^{\prime}_1}^2  + {g^{\prime}_2}^2 } \ , \qquad  M_{W^{\prime}} = \tilde{v}
 \sqrt{{g_1}^2  + {g_2}^2 } 
 \end{eqnarray}
 and four massless ones: 
 \begin{equation}
 B= \cos \beta \, B_2 + \sin \beta \, B_1 \ , \qquad {W}^a =    \cos \omega \,  W_2^a + \sin \omega \, W_1^a \ , 
 \end{equation}
which we identify with the SM states. The two mixing angles are related to the gauge couplings as follow\begin{eqnarray}
\cos \beta = \frac{g^{\prime}_1}{\sqrt{{g^{\prime}_1}^2  + {g^{\prime}_2}^2 }}  ~~~~~~~~~~~~  \sin \beta = \frac{g^{\prime}_2}{\sqrt{{g^{\prime}_1}^2  + {g^{\prime}_2}^2 }}  \ ,
\end{eqnarray} and 
\begin{eqnarray}
\cos \omega = \frac{g_1}{\sqrt{{g_1}^2  + {g_2}^2 }}   ~~~~~~~~~~~~ \sin \omega = \frac{g_2}{\sqrt{{g_1}^2  + {g_2}^2 }}  \ .
\end{eqnarray}
The covariant derivative acting on purely SM fields before breaking the extended symmetry is:
\begin{eqnarray}
D = \partial - i g_2\, W_2^a T^a - i g_2^{\prime}Y B_2  \ ,  \qquad {\rm SM-particles}
\end{eqnarray}
with $T^a$ the $SU_2(2)$ generators and $Y$ the generator of $U_2(1)$ for a generic SM particle. After symmetry breaking the covariant derivative becomes: 
 \begin{eqnarray}
D = \partial - i g\, W^a T^a - i g^{\prime}Y B +   i g \tan \omega\, {W^{\prime}}^a T^a + i g^{\prime}\tan \beta\,Y B^{\prime} \ ,  \quad {\rm SM-particles} 
\end{eqnarray}
with $g=g_2  \cos \omega$ and $g^{\prime}=g^{\prime}_2 \cos\beta$ the SM couplings. 

{}For the new leptons and techniquarks (Q) the electroweak covariant derivative, after symmetry breaking, and in terms of the gauge bosons mass eigenstates is:
\begin{eqnarray}
D = \partial - i g\, W^a T^a - i g^{\prime}Y B -  i \frac{g}{ \tan\omega}\, {W^{\prime}}^a T^a - i \frac{g^{\prime}}{\tan \beta}\,Y B^{\prime} \ ,  \quad {\rm New-Leptons~\&~Q} 
\end{eqnarray}
This shows that the new leptons have the same coupling of the SM leptons to the SM gauge fields but will couple differently to the new gauge bosons. In the limit in which $\tilde{v}$ is much larger than the electroweak symmetry we recover the previous model analysis. The constraints and phenomenological consequences of the model we have just introduced will be investigated elsewhere.

%%%%%%%%%%%%%%%%%%%%%%%%%%%%%%%%%%%%%%%%%%%%%%%%%%%%%%%%%%%%
\section{Conclusions and outlook}

We have investigated the LHC phenomenology of a fourth family of leptons whose appearance is justified by the presence of a new strongly coupled sector. The latter is responsible for the breaking of the electroweak symmetry dynamically. We have chosen as template for our analysis MWT.  In this way the fourth family as well as the Higgs sector of the SM are natural theories. We analyzed a general heavy neutrino mass structure with and without mixing with the SM families. We have then shown that the interplay of heavy neutrinos and composite vector mesons can lead to striking signatures at the LHC.
Finally we introduced a model uniting the fourth lepton family and the technifermion sector at higher energies.

\subsection*{Note added}
While this work was being completed the paper \cite{Antipin:2009ks} appeared in which some of the ideas considered  here were also discussed. 

%\section{Conclusions}
%
%In this paper we have emphasized how heavy leptons naturally appear in recent walking Technicolor models. We have considered the interplay between the heavy leptons and the composite scalar and vector Technicolor states, in different scenarios, including a fairly general scheme for mixing with the SM states. 

%An interesting possibility in the walking Technicolor sector, as opposed to QCD, is the possibility of a light and narrow axial resonance (While in our parametrization the axial resonance is even lighter than the lightest vector resonance, the main interesting point is just that the axial be light and narrow).    
%We found an interesting interplay between the axial resonance and a new heavy majorana neutrino state, manifesting itself in distinct LHC signatures for invisible composite higgs decays (if the heavy majorana neutrino is stable) or in same sign di-lepton production (if the heavy Majorana neutrino mixes with the SM leptons). 

%In the future it would be interesting to consider in more detail the $n.1$ model as well as the effect of the extended scalar sector.

%%%%%%%%%%%%%%%%%%%%%%%%%%%%%%%%%%%%%%%%%%%%%%%%%%%%%%%%%

%%%%%%%%%%%%%%%%%%%%%%%%%%%%%%%%%%%%%%%%%%%%%%%%%%%%%%%%%%%%%%%%%%%%%%%%%%%%%%%%%%%%%%%%

\appendix

\section{Full mixing}
\label{sec:full mixing}
The mixing between the neutrinos of flavor $\zeta$ and the 3 SM neutrinos is described by the $5\times 5$
mass matrix ${\cal M}$: 
\begin{equation}
-{\cal L}= \frac{1}{2} ( \begin{array}{ccccc} 
 \overline{\nu_{e L}}&\overline{\nu_{\mu L}}&\overline{\nu_{\tau L}}& \overline{\nu_{\zeta L}} & \overline{(\nu_{\zeta R})^c}  \end{array} )
\left( \begin{array}{ccccc}  
{\cal O}(eV) & {\cal O}(eV) &{\cal O}(eV) & {\cal O}(eV) & m'_e \cr  
{\cal O}(eV) & {\cal O}(eV) &{\cal O}(eV) & {\cal O}(eV) & m'_\mu \cr  
{\cal O}(eV) & {\cal O}(eV) &{\cal O}(eV) & {\cal O}(eV) & m'_\tau \cr  
{\cal O}(eV) & {\cal O}(eV) &{\cal O}(eV) & {\cal O}(eV) & m_D \cr 
m'_e & m'_\mu& m'_\tau & m_D &  m_R \end{array} \right)  
\left( \begin{array}{c} (\nu_{e L})^c \cr(\nu_{\mu L})^c \cr(\nu_{\tau L})^c \cr (\nu_{\zeta L})^c \cr \nu_{\zeta R}  \end{array} \right) + h.c.~~.
\label{}
\end{equation}
Barring unnatural tunings and up to corrections to its mixings of ${\cal O}(eV/M_{1,2})$,
the unitary matrix $V$ 
\begin{equation}
\left( \begin{array}{c}
\nu_{e L} \cr\nu_{\mu L} \cr\nu_{\tau L} \cr  \nu_{\zeta L} \cr  (\nu_{\zeta R})^c  \end{array} \right) = V
\left( \begin{array}{c} P_L N_e \cr  P_L N_\mu \cr P_L N_\tau \cr P_L N_1 \cr P_L N_2  \end{array} \right) ~~,~~
V=\left( \begin{array}{ccccc} 
c_e & -s_e s_\mu & -s_e c_\mu s_\tau &  s_e c_\mu c_\tau & 0 \cr
0 & c_\mu & - s_\mu s_\tau & - s_\mu c_\tau & 0 \cr
0 & 0 & c_\tau & -s_\tau & 0 \cr 
i c s_e & i c c_e s_\mu & i c c_e c_\mu s_\tau & i c c_e c_\mu c_\tau & -i s \cr 
s s_e & s c_e s_\mu & s c_e c_\mu s_\tau & s c_e c_\mu c_\tau & c 
\end{array} \right)~,
\label{}
\end{equation}
where 
\bea
t_\tau^2 &=& \frac{{m'}_\tau^2}{m_D^2} ~~~, ~~~~~~ t_\mu^2 = \frac{{m'}_\mu^2}{ m_D^2 + {m'}_\tau^2} ~~~,~~~~~
t_e^2 = \frac{{m'}_e^2}{m_D^2 + {m'}_\tau^2+{m'}_\mu^2} ~~,\\
&~&~~~~~~\tan( 2 \theta)= 2 ~\frac{{m'}_D}{m_R} ~~,~~~{m'}_D^2=m_D^2+{m'}_\tau^2+{m'}_\mu^2+{m'}_e^2\no,
\label{}
\eea
diagonalises the lower sector of the $5\times 5$ mass matrix ${\cal M}$, namely
\beq
V {\cal M} V^\dagger =  \left( \begin{array}{ccc}  
 m^{eff}_{3\times 3} & 0 & 0 \cr  
 0 & M_1 & 0 \cr 
0  &  0 &  M_2 \end{array} \right)  
\label{matnu}
\eeq
where the elements of $m^{eff}_{3\times 3}$ are naturlly ${\cal O}(eV)$ and
\beq
M_1=\frac{m_R}{2} \left( \sqrt{1+4 \frac{{m'_D}^2}{m_R^2}} -1 \right) ~~,~~~~
M_2=\frac{m_R}{2} \left( \sqrt{1+4 \frac{{m'_D}^2}{m_R^2}} +1 \right) ~.
\eeq
At this stage, the matrix in eq.(\ref{matnu}) can be fully diagonalised with a further unitary matrix,
which can be identified with the MNS mixing matrix, acting only on the upper $3\times 3$ block:
\beq
U^\dagger m^{eff}_{3\times 3} U^* = {\rm diag}(m_1,m_2,m_3)~~.
\eeq

\section{MWT effective Lagrangian}
\label{effective}
Following  
Ref.~\cite{Foadi:2007ue} and \cite{Appelquist:1999dq} the MWT effective Lagrangian containing spin one and spin zero states reads:
\begin{eqnarray}
{\cal L}_{\rm boson}&=&-\frac{1}{2}{\rm Tr}\left[\widetilde{W}_{\mu\nu}\widetilde{W}^{\mu\nu}\right]
-\frac{1}{4}\widetilde{B}_{\mu\nu}\widetilde{B}^{\mu\nu}
-\frac{1}{2}{\rm Tr}\left[F_{{\rm L}\mu\nu} F_{\rm L}^{\mu\nu}+F_{{\rm R}\mu\nu} F_{\rm R}^{\mu\nu}\right] \nonumber \\
&+& m^2\ {\rm Tr}\left[C_{{\rm L}\mu}^2+C_{{\rm R}\mu}^2\right]
+\frac{1}{2}{\rm Tr}\left[D_\mu M D^\mu M^\dagger\right]
-\tilde{g^2}\ r_2\ {\rm Tr}\left[C_{{\rm L}\mu} M C_{\rm R}^\mu M^\dagger\right] \nonumber \\
&-&\frac{i\ \tilde{g}\ r_3}{4}{\rm Tr}\left[C_{{\rm L}\mu}\left(M D^\mu M^\dagger-D^\mu M M^\dagger\right)
+ C_{{\rm R}\mu}\left(M^\dagger D^\mu M-D^\mu M^\dagger M\right) \right] \nonumber \\
&+&\frac{\tilde{g}^2 s}{4} {\rm Tr}\left[C_{{\rm L}\mu}^2+C_{{\rm R}\mu}^2\right] {\rm Tr}\left[M M^\dagger\right]
+\frac{\mu^2}{2} {\rm Tr}\left[M M^\dagger\right]-\frac{\lambda}{4}{\rm Tr}\left[M M^\dagger\right]^2
\label{eq:boson}
\end{eqnarray}
where $\widetilde{W}_{\mu\nu}$ and $\widetilde{B}_{\mu\nu}$ are the ordinary electroweak field strength tensors, 
$F_{{\rm L/R}\mu\nu}$ are the field strength tensors associated to the vector meson fields $A_{\rm L/R\mu}$. 
$V=\frac{1}{\sqrt{2}}(A_L + A_R) , A=\frac{1}{\sqrt{2}}(A_L - A_R)$ are the mass eigenstates in the $g,g'\to 0$ 
limit (the analogs of the QCD $\rho$ and $a_0$ vector and axial-vector mesons).  
The $C_{{\rm L}\mu}$ and $C_{{\rm R}\mu}$ fields are
\begin{eqnarray}
C_{{\rm L}\mu}\equiv A_{{\rm L}\mu}-\frac{g}{\tilde{g}}\widetilde{W_\mu}\ , \quad
C_{{\rm R}\mu}\equiv A_{{\rm R}\mu}-\frac{g^\prime}{\tilde{g}}\widetilde{B_\mu}\ .
\end{eqnarray}
The 2$\times$2 matrix $M$ is
\begin{eqnarray}
M=\frac{1}{\sqrt{2}}\left[v+H+2\ i\ \pi^a\ T^a\right]\ ,\quad\quad  a=1,2,3
\end{eqnarray}
where $\pi^a$ are the Goldstone bosons produced in the chiral symmetry breaking, $v=\mu/\sqrt{\lambda}$ is the 
corresponding VEV, $H$ is the composite Higgs, and $T^a=\sigma^a/2$, where $\sigma^a$ are the Pauli matrices. 
The covariant derivative is
\begin{eqnarray}
D_\mu M&=&\partial_\mu M -i\ g\ \widetilde{W}_\mu^a\ T^a M + i\ g^\prime \ M\ \widetilde{B}_\mu\ T^3\ . 
\label{Eq:covderivM}
\end{eqnarray}
When $M$ acquires its VEV, the Lagrangian of Eq.~(\ref{eq:boson}) contains mixing matrices for the spin one fields. 
The mass eigenstates are the ordinary SM bosons, and two triplets of heavy mesons, of which the lighter (heavier) 
ones are denoted by $R_1^\pm$ ($R_2^\pm$) and $R_1^0$ ($R_2^0$). Besides the new heavy leptons, these heavy mesons 
are the only new particles, at low energy, relative to the SM.

The main input parameters in our implementation of the (N)MWT models are 
\bea
\label{TCparameters}
M_Z \ , G_F \ , e \ , S \ , M_A \ , \tilde{g} \ , 
\eea
where the latter three are parameters of the Technicolor sector per se. The S parameter we fix to be 
$S=0.1,0.3$ MWT and NMWT models respectively from an estimate based on the underlying dynamics. We then have still the overall mass scale and coupling for the heavy vectors $M_A , \tilde{g}$  to scan over. The mass mixing in the neutral sector between the gauge eigenvectors $B, \widetilde{W}, A, V$ is then diagonalized by a $4\times 4$ orthogonal matrix with entries $N_{ij}(\tilde{g}, M_A)$ 
\begin{eqnarray}
B&=& N_{11} A + N_{12} Z + N_{13} R_1^0 + N_{14} R_2^0 \nonumber \\
\widetilde{W}^3&=& N_{21} A + N_{22} Z + N_{23} R_1^0 + N_{24} R_2^0 \nonumber \\
A^3&=& N_{31} A + N_{32} Z + N_{33} R_1^0 + N_{34} R_2^0 \nonumber \\
V^3&=& N_{41} A + N_{42} Z + N_{43} R_1^0 + N_{44} R_2^0 \ .
\label{vector mixing}
\end{eqnarray}
In the limit $\tilde{g}\to \infty$ we have $N_{33} , N_{44} \to 1$ while $N_{ij} , i,j=1,2$ take their SM 
values and all other coefficients go to zero. 
Upon diagonalization, we organize the mass eigenstates $R_i$ such that $M_{R_1}< M_{R_2}$. 
The first Weinberg Sum Rule is imposed over the massive vector states. The second Weinberg sum rule 
is assumed to be modified due to the walking behaviour of the underlying gauge theory. 

Consequently the mass spectrum and the mixing coefficients are fixed in terms $M_A , \tilde{g}$ 
\cite{Foadi:2007ue,Belyaev:2008yj} . We reproduce the plots from the latter reference below:
\begin{figure}
{
\includegraphics[width=0.45\textwidth]{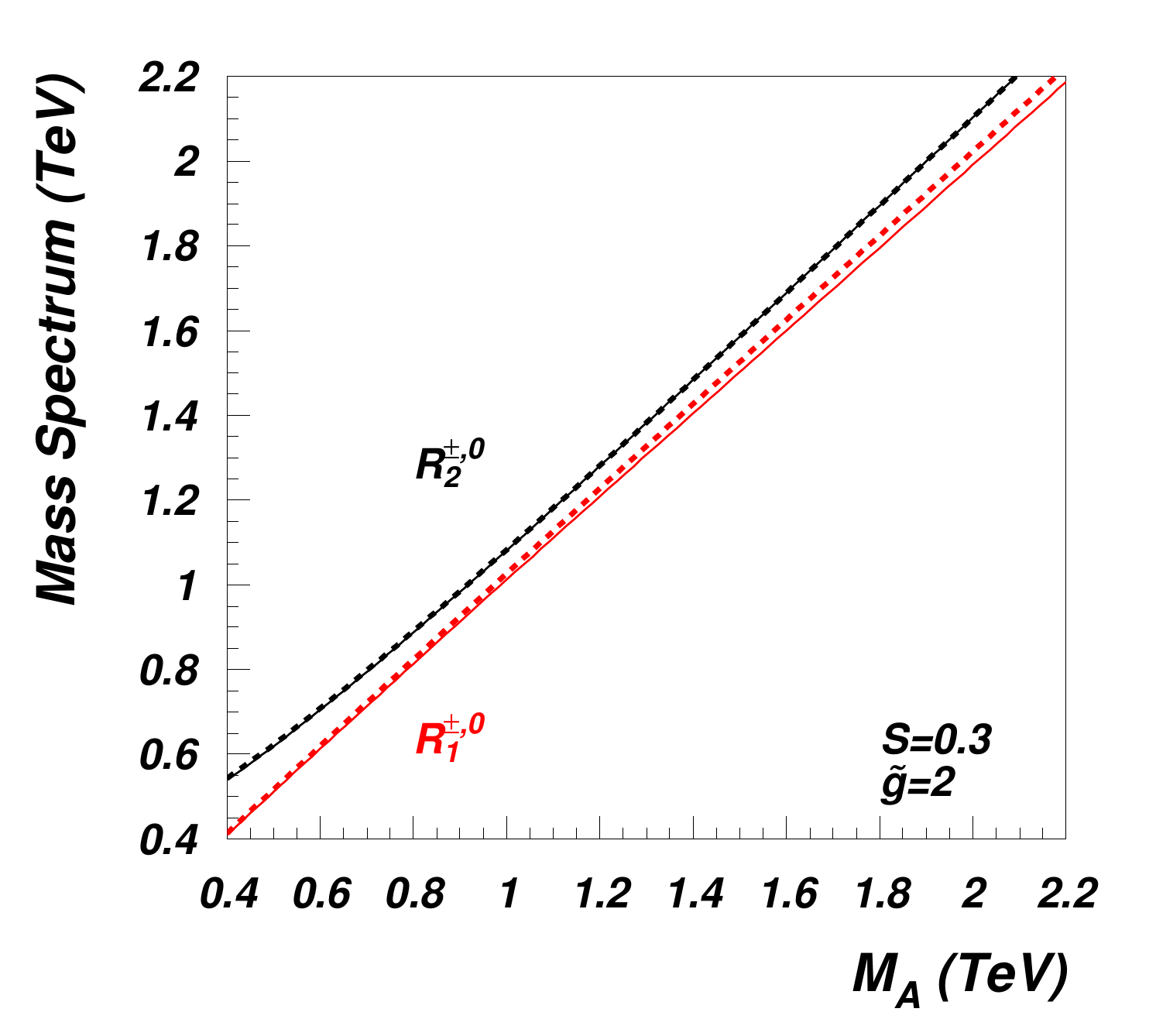}
\includegraphics[width=0.45\textwidth]{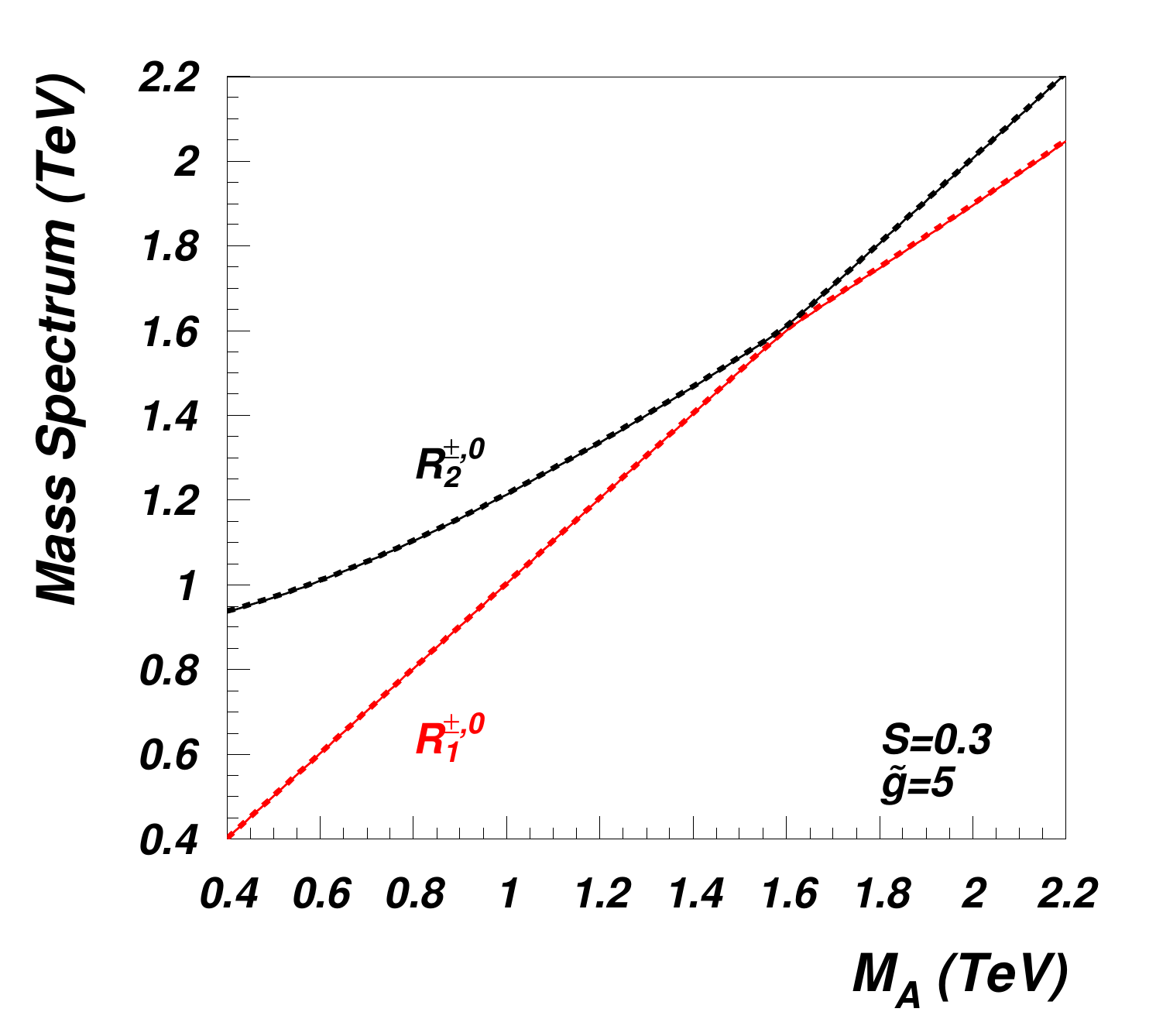}%
} 
 \caption{The mass spectrum of  the $M_{R^{\pm,0}_{1,2}}$
        vector mesons versus $M_A$
	for $\tilde{g}=2$ (left)
	and $\tilde{g}=5$ (right).
	The masses of the charged vector mesons are denoted by solid lines,
	while the masses of the neutral mesons are denoted by dashed lines.
	\label{fig:spectrum}}
\end{figure}
For $\tilde{g}=5$ the inversion point around 1.6 TeV below which $R_1$ is (mostly) axial while above it 
$R_2$ is mostly axial (as in QCD) is visible. 
For $\tilde{g}=2$ where the mixing due to electroweak effects the inversion point is not visible but still 
occurs in the pure TC sector for $g,g'=0$ and still affects processes as detailed in 
\cite{Foadi:2007ue,Belyaev:2008yj}.
For $S=0.1$ the inversion point happens at a considerably larger value of $M_A \sim 2.5$ TeV.

In this study we will focus on the parameter values $S=0.1,M_A=500$ and we will take two values of the 
coupling $\tilde{g}=2,5$. The physical parameters with these choices are 
\bea
M_{R_1}= 503 GeV \ , \  M_{R_2}= 996 GeV \ , \Gamma[R_1]=0.2 GeV\ , \Gamma[R_2]=296 GeV
\nn
M_{R_1}= 518 GeV \ , \  M_{R_2}= 591 GeV \ , \Gamma[R_1]=1 GeV\ , \Gamma[R_2]=4 GeV
\label{Eq:Masses and widths of R12}
\eea

\section{EW parameters}
\label{gates}
The S,T and U parameters \cite{Peskin:1991sw} in our setup, for $M_\zeta , M_N > M_Z$ for a lepton doublet and additional right-handed singlet as computed in \cite{Gates:1991uu}. We keep the hypercharge dependence explicit. 
\bea
S &=& \frac{1}{6\pi}\big\{-Y \big( c_\theta^2 \ln \frac{M_1^2}{M_\zeta^2} + s_\theta^2 \ln \frac{M_2^2}{M_\zeta^2}\big) +\frac{3}{2} 
\nn
&-& s_\theta^2 c_\theta^2 \big(\frac{8}{3}+f_1(M_1,M_2)-f_2(M_1,M_2)\ln \frac{M_1^2}{M_2^2} \big) \big\} 
\nn 
T &=& \frac{1}{16\pi s_W^2 c_W^2 M_Z^2}\times 
\nn
&\phantom{1}& \big\{ 
c_\theta^2 \big( M_1^2+M_\zeta^2-\frac{2 M_1^2 M_\zeta^2}{M_1^2-M_\zeta^2} \ln \frac{M_1^2}{M_\zeta^2}     \big) 
+
s_\theta^2 \big( M_2^2+M_\zeta^2-\frac{2 M_2^2 M_\zeta^2}{M_2^2-M_\zeta^2} \ln \frac{M_2^2}{M_\zeta^2}    
\big) 
\nn
&-& s_\theta^2c_\theta^2\big(M_1^2+M_2^2-4 M_1 M_2+2\frac{M_1^3 M_2-M_1^2M_2^2+M_1 M_2^3}{M_1^2-M_2^2} \ln \frac{M_1^2}{M_2^2}
\big)
\big\}
\eea

The functions $f_1$ and $f_2$ are given by 
\bea 
f_1(M_1,M_2)&=&\frac{3M_1 M_2^3+3M_1^3M_2-4M_1^2M_2^2}{(M_1^2-M_2^2)^2}
\nn
f_2(M_1,M_2)&=&\frac{M_1^6-3M_1^4M_2^2+6M_1^3M_2^3-3M_1^2M_2^4+M_2^6}{(M_1^2-M_2^2)^3}
\eea

\section{Decay widths of the heavy leptons}
\label{decays}
\underline{\bf Decays of the heavy leptons}
The partial widths of $N_2$ before including any mixing with the SM leptons are given by.  
\bea
\Gamma[N_2\to W^\pm \zeta^\mp] &=& \frac{e^2 s^2c'^2}{64\pi s_w^2 M_2^3}
\sqrt{M_2^4-2 M_2^2 M_+^2+M_-^4}
\\ \nonumber 
&\times & [(M_2^2 + M_-^2) +\frac{(M_2^2 - M_-^2)(M_2^2 - M_+^2)}{M_W^2}]
% \\ \nonumber 
%&\to & \frac{e^2 |V_{N_i\ell}|^2}{64\pi s_w^2}\ \frac{M_i^2}{M_W^2}
%(1-\frac{M_W^2}{M_i^2})(1+\frac{M_W^2}{M_i^2}-2\frac{M_W^4}{M_i^4})
\\ \nonumber
\Gamma[N_1\to W^\pm \zeta^\mp] &=& \frac{e^2 c^2c'^2}{64\pi s_w^2 M_1^3}
\sqrt{M_1^4-2 M_1^2 M_+^2+M_-^4}
\\ \nonumber 
&\times & [(M_1^2 + M_-^2) +\frac{(M_1^2 - M_-^2)(M_1^2 - M_+^2)}{M_W^2}]
\\ \nonumber
\Gamma[\zeta^\pm \to W^\pm N_1] &=& \frac{e^2 c^2c'^2}{64\pi s_w^2 M_\zeta^3}
\sqrt{M_\zeta^4-2 M_\zeta^2 M_+^{'2}+M_-^{'4}}
\\ \nonumber 
&\times & [(M_\zeta^2 + M_-^{'2}) +\frac{(M_\zeta^2 - M_-^{'2})(M_\zeta^2 - M_+^{'2})}{M_W^2}]
\\ \nonumber
\Gamma[\zeta^\pm\to W^\pm N_2] &=& \frac{e^2 s^2c'^2}{64\pi s_w^2 M_\zeta^3}
\sqrt{M_\zeta^4-2 M_\zeta^2 M_+^{'''2}+M_-^{'''4}}
\\ \nonumber 
&\times & [(M_\zeta^2 + M_-^{'''2}) +\frac{(M_\zeta^2 - M_-^{'''2})(M_\zeta^2 - M_+^{'''2})}{M_W^2}]
\\ \nonumber
\Gamma[N_2\to Z N_1] &=& \frac{e^2 c^2 s^2}{64\pi s_w^2 c_w^2 M_2^3}
\sqrt{M_2^4-2 M_2^2 M_+^{'2}+M_-^{'4}}
\\ \nonumber 
&\times & [(M_2^2 + M_-^{'2}) +\frac{(M_2^2 - M_-^{'2})(M_2^2 - M_+^{'2})}{M_Z^2}-6M_1M_2]
\\ \nonumber 
\Gamma[N_2\to H N_1] &=& \frac{e^2 (c^2 -s^2)^2 c'^2 m_D^2}{64\pi s_w^2 M_W^2 M_2^3}
\sqrt{M_2^4-2 M_2^2 M_+^{''2}+M_-^{''4}}
\\ \nonumber 
&\times & [(M_2^2 + M_-^{''2}) -2M_1M_2]
\eea
where $M_{\pm}^2=M_\zeta^2\pm M_W^2$, $M_{\pm}^{'2}=M_1\pm M_W^2$, $M_{\pm}^{''2}=M_1\pm M_H^2$, $M_{\pm}^{'''2}=M_2\pm M_W^2$. 
The additional decays of $N_{1,2}$ induced by the mixing are  
\bea
\Gamma[N_2\to W^\pm \ell^\mp] &=& \frac{e^2 s^2 s'^2 M_2^3}{64\pi s_w^2 M_W^2}(1-\frac{M_W^2}{M_2^2})(1+\frac{M_W^2}{M_2^2}-2\frac{M_W^4}{M_2^4})
\nn \\
\Gamma[N_1\to W^\pm \ell^\mp] &=& \frac{e^2 c^2 s'^2 M_1^3}{64\pi s_w^2 M_W^2}(1-\frac{M_W^2}{M_1^2})(1+\frac{M_W^2}{M_1^2}-2\frac{M_W^4}{M_1^4})
\nn \\
\Gamma[\zeta^\pm \to W^\pm N_0] &=& \frac{e^2 s'^2 M_\zeta^3}{64\pi s_w^2 M_W^2}(1-\frac{M_W^2}{M_\zeta^2})(1+\frac{M_W^2}{M_\zeta^2}-2\frac{M_W^4}{M_\zeta^4})
\nn \\
\Gamma[N_1\to H N_0] &=& \frac{e^2 s^2 s'^2 m_D^2}{64\pi s_w^2 M_W^2 M_1^3}
(M_1^2-M_H^2)^2 
\eea
The partial widths of the Higgs into the 4th family are given by 
\bea 
\Gamma[H\to \zeta^+ \zeta^-] &=& \frac{ e^2 m_\zeta^2}{32\pi s_w^2 M_W^2}(1-\frac{4 m_\zeta^2}{M_H^2})
\sqrt{1-\frac{4 m_\zeta^2}{M_H^2}}
\\ \nonumber 
\Gamma[H\to N_i N_i] &=& \frac{ e^2 c^2 s^2 m_D^2}{16\pi s_w^2 M_W^2}(1-\frac{4M_i^2}{M_H^2})
\sqrt{1-\frac{4 M_i^2}{M_H^2}}
\eea

\end{document}